\documentclass[apj]{emulateapj}

\usepackage{textcomp}
\usepackage{bm,color}
\usepackage{verbatim}
\usepackage{amsmath}
\usepackage{ulem}

\newcommand{\rotval}{-1.08}
\newcommand{\rotuncert}{0.20}

\newcommand{\ABB}{$A_{BB}$}
\newcommand{\ABBmeas}{$A_{BB} = 1.12 \pm 0.61 ({\rm stat}) ^{+0.04}_{-0.12}({\rm sys}) \pm 0.07 ({\rm multi})$}
\newcommand{\ptelcdm}{42\%}
\newcommand{\rejectnull}{97.1\%}

\newcommand{\rejectnullcombinedsigma}{4.7$\sigma$} %sqrt(4.2**2+2.2**2) = 4.74130783645187925e+00

\newcommand{\pp}{\parallel/\perp}

\newcommand{\pb}{\textsc{Polarbear}}
\newcommand{\Pb}{\textsc{Polarbear}}
\newcommand{\uKsq}{$\mu \mbox{K}^2$}
\renewcommand{\textdegree}{$^\circ$}

\newcommand{\rahrs}{$^\mathrm{h}$}
\newcommand{\ramin}{$^\mathrm{m}$}
\newcommand{\rasec}{$^\mathrm{s}$}

\newcommand{\secref}{Section}
\newcommand{\Secref}{Section}
\newcommand{\figref}{Figure}
\newcommand{\tabref}{Table}
\newcommand{\Figref}{Figure}
\newcommand{\Tabref}{Table}

\newcommand{\secrefs}{Sections}

\newcommand{\figrefs}{Figures}

\newcommand{\mc}{Monte Carlo}

\newcommand{\apex}{{\sc apex}}

\newcommand{\spt}{{\sc spt}}
\newcommand{\sptpol}{{\sc SPTpol}}

\newcommand{\planck}{Planck}

\newcommand{\sqdeg}{$\deg^2$}

\newcommand{\clbb}{$C_\ell^{BB}$}
\newcommand{\cleb}{$C_\ell^{EB}$}
\newcommand{\cltt}{$C_\ell^{TT}$}
\newcommand{\clte}{$C_\ell^{TE}$}
\newcommand{\cltb}{$C_\ell^{TB}$}
\newcommand{\clee}{$C_\ell^{EE}$}

\newcommand{\dlbb}{$\ell \left ( \ell + 1 \right ) C_\ell^{BB} / 2\pi$}

\newcommand{\taua}{Tau A}
\newcommand{\cena}{Cen A}
\newcommand{\Mll}{$M_{\ell\ell^\prime}$}

\newcommand{\quiet}{{\sc quiet}}

\newcommand{\quadexp}{{QUaD}}

\newcommand{\wmap}{{\sc wmap}}

\newcommand{\npix}{637}
\newcommand{\nbolo}{1,274}

\newcommand{\beq}{\begin{equation}}
\newcommand{\eeq}{\end{equation}}
\newcommand{\bea}{\begin{eqnarray}}
\newcommand{\eea}{\end{eqnarray}}
\newcommand{\lcdm}{$\Lambda$CDM}

\newcommand{\microKrtsec}{$\mu{\mbox{K}}\sqrt{\mbox{s}}$}

\newcommand{\uK}{$\mu\mbox{K}$}

\begin{document}

\title{A Measurement of the Cosmic Microwave Background $B$-Mode Polarization Power Spectrum at Sub-Degree Scales with POLARBEAR}

\shorttitle{\pb\ CMB $B$-Mode Power Spectrum}

\author{The \pb\ Collaboration: 
P.A.R. Ade\altaffilmark{29},
Y. Akiba\altaffilmark{32},
A.E. Anthony\altaffilmark{2,5},
K. Arnold\altaffilmark{14},
M. Atlas\altaffilmark{14},
D. Barron\altaffilmark{14},
D. Boettger\altaffilmark{14},
J. Borrill\altaffilmark{3,31},
S. Chapman\altaffilmark{9},
Y. Chinone\altaffilmark{17,13},
M. Dobbs\altaffilmark{25},
T. Elleflot\altaffilmark{14},
J. Errard\altaffilmark{31,3},
G. Fabbian\altaffilmark{1,18},
C. Feng\altaffilmark{14},
D. Flanigan\altaffilmark{13,10},
A. Gilbert\altaffilmark{25},
W. Grainger\altaffilmark{28},
N.W. Halverson\altaffilmark{2,5,15},
M. Hasegawa\altaffilmark{17,32},
K. Hattori\altaffilmark{17},
M. Hazumi\altaffilmark{17,32,20},
W.L. Holzapfel\altaffilmark{13},
Y. Hori\altaffilmark{17},
J. Howard\altaffilmark{13,16},
P. Hyland\altaffilmark{24},
Y. Inoue\altaffilmark{32},
G.C. Jaehnig\altaffilmark{2,15},
A.H. Jaffe\altaffilmark{11},
B. Keating\altaffilmark{14},
Z. Kermish\altaffilmark{12},
R. Keskitalo\altaffilmark{3},
T. Kisner\altaffilmark{3,31},
M. Le Jeune\altaffilmark{1},
A.T. Lee\altaffilmark{13,27},
E.M. Leitch\altaffilmark{4,19},
E. Linder\altaffilmark{27},
M. Lungu\altaffilmark{13,8},
F. Matsuda\altaffilmark{14},
T. Matsumura\altaffilmark{17},
X. Meng\altaffilmark{13},
N.J. Miller\altaffilmark{22},
H. Morii\altaffilmark{17},
S. Moyerman\altaffilmark{14},
M.J. Myers\altaffilmark{13},
M. Navaroli\altaffilmark{14},
H. Nishino\altaffilmark{20},
A. Orlando\altaffilmark{14},
H. Paar\altaffilmark{14},
J. Peloton\altaffilmark{1},
D. Poletti\altaffilmark{1},
E. Quealy\altaffilmark{13,26},
G. Rebeiz\altaffilmark{6},
C.L. Reichardt\altaffilmark{13},
P.L. Richards\altaffilmark{13},
C. Ross\altaffilmark{9},
I. Schanning\altaffilmark{14},
D.E. Schenck\altaffilmark{2,5},
B.D. Sherwin\altaffilmark{13,21},
A. Shimizu\altaffilmark{32},
C. Shimmin\altaffilmark{13,7},
M. Shimon\altaffilmark{30,14},
P. Siritanasak\altaffilmark{14},
G. Smecher\altaffilmark{33},
H. Spieler\altaffilmark{27},
N. Stebor\altaffilmark{14},
B. Steinbach\altaffilmark{13},
R. Stompor\altaffilmark{1},
A. Suzuki\altaffilmark{13},
S. Takakura\altaffilmark{23,17},
T. Tomaru\altaffilmark{17},
B. Wilson\altaffilmark{14},
A. Yadav\altaffilmark{14},
O. Zahn\altaffilmark{27}
}
\altaffiltext{1}{AstroParticule et Cosmologie, Univ Paris Diderot, CNRS/IN2P3, CEA/Irfu, Obs de Paris, Sorbonne Paris Cit\'e, France}
\altaffiltext{2}{Center for Astrophysics and Space Astronomy, University of Colorado, Boulder, CO 80309, USA}
\altaffiltext{3}{Computational Cosmology Center, Lawrence Berkeley National Laboratory, Berkeley, CA 94720, USA}
\altaffiltext{4}{Department of Astronomy and Astrophysics, University of Chicago, Chicago, IL 60637, USA}
\altaffiltext{5}{Department of Astrophysical and Planetary Sciences, University of Colorado, Boulder, CO 80309, USA}
\altaffiltext{6}{Department of Electrical and Computer Engineering, University of California, San Diego, CA 92093, USA}
\altaffiltext{7}{Department of Physics and Astronomy, University of California, Irvine, CA 92697-4575, USA}
\altaffiltext{8}{Department of Physics and Astronomy, University of Pennsylvania, Philadelphia, PA 19104-6396, USA}
\altaffiltext{9}{Department of Physics and Atmospheric Science, Dalhousie University, Halifax, NS, B3H 4R2, Canada}
\altaffiltext{10}{Department of Physics, Columbia University, New York, NY 10027, USA}
\altaffiltext{11}{Department of Physics, Imperial College London, London SW7 2AZ, United Kingdom}
\altaffiltext{12}{Department of Physics, Princeton University, Princeton, NJ 08544, USA}
\altaffiltext{13}{Department of Physics, University of California, Berkeley, CA 94720, USA}
\altaffiltext{14}{Department of Physics, University of California, San Diego, CA 92093-0424, USA}
\altaffiltext{15}{Department of Physics, University of Colorado, Boulder, CO 80309, USA}
\altaffiltext{16}{Department of Physics, University of Oxford, Oxford OX1 2JD, United Kingdom}
\altaffiltext{17}{High Energy Accelerator Research Organization (KEK), Tsukuba, Ibaraki 305-0801, Japan}
\altaffiltext{18}{International School for Advanced Studies (SISSA), Trieste 34014, Italy}
\altaffiltext{19}{Kavli Institute for Cosmological Physics, University of Chicago, Chicago, IL 60637, USA}
\altaffiltext{20}{Kavli Institute for the Physics and Mathematics of the Universe (WPI), Todai Institutes for Advanced Study, The University of Tokyo, Kashiwa, Chiba 277-8583, Japan}
\altaffiltext{21}{Miller Institute for Basic Research in Science, University of California, Berkeley, CA 94720, USA}
\altaffiltext{22}{Observational Cosmology Laboratory, Code 665, NASA Goddard Space Flight Center, Greenbelt, MD 20771, USA}
\altaffiltext{23}{Osaka University, Toyonaka, Osaka 560-0043, Japan}
\altaffiltext{24}{Physics Department, Austin College, Sherman, TX 75090, USA}
\altaffiltext{25}{Physics Department, McGill University, Montreal, QC H3A 0G4, Canada}
\altaffiltext{26}{Physics Department, Napa Valley College, Napa, CA 94558, USA}
\altaffiltext{27}{Physics Division, Lawrence Berkeley National Laboratory, Berkeley, CA 94720, USA}
\altaffiltext{28}{Rutherford Appleton Laboratory, STFC, Swindon, SN2 1SZ, United Kingdom}
\altaffiltext{29}{School of Physics and Astronomy, Cardiff University, Cardiff CF10 3XQ, United Kingdom}
\altaffiltext{30}{School of Physics and Astronomy, Tel Aviv University, Tel Aviv 69978, Israel}
\altaffiltext{31}{Space Sciences Laboratory, University of California, Berkeley, CA 94720, USA}
\altaffiltext{32}{The Graduate University for Advanced Studies, Hayama, Miura District, Kanagawa 240-0115, Japan}
\altaffiltext{33}{Three-Speed Logic, Inc., Vancouver, B.C., V6A 2J8, Canada}

\begin{abstract}
We report a measurement of the $B$-mode polarization power spectrum in the cosmic microwave background (CMB) using the \pb\ experiment in Chile.
The faint $B$-mode polarization signature carries information about the Universe's entire history of gravitational structure formation, and the cosmic inflation that may have occurred in the very early Universe.  
Our measurement covers the angular multipole range $500 < \ell < 2100$ and is based on observations of an effective sky area of 25\,\sqdeg\ with 3.5\arcmin\ resolution at 150\,GHz.
On these angular scales, gravitational lensing of the CMB by intervening structure in the Universe is expected to be the dominant source of $B$-mode polarization.
Including both systematic and statistical uncertainties, the hypothesis of no $B$-mode polarization power from gravitational lensing is rejected at \rejectnull\ confidence.
The band powers are consistent with the standard cosmological model. 
Fitting a single lensing amplitude parameter \ABB\ to the measured band powers, \ABBmeas, where $A_{BB} = 1$ is the fiducial \wmap-9 \lcdm\ value. 
In this expression, ``stat'' refers to the statistical uncertainty, ``sys'' to the systematic uncertainty associated with possible biases from the instrument and astrophysical foregrounds, and ``multi'' to the calibration uncertainties that have a multiplicative effect on the measured amplitude $A_{BB}$. 
\end{abstract}


\section{Introduction}

\setcounter{footnote}{0}

The cosmic microwave background (CMB) radiation is emitted from the primordial plasma in the early universe when stable hydrogen first forms at a redshift of $z = 1091$ \citep{2013ApJS..208...20B}. 
The scalar density fluctuations present in the primordial plasma at that time, which are the seeds for later structure formation, create both CMB intensity and polarization anisotropies. 
These scalar fluctuations can only create polarization patterns of even spatial parity, referred to as $E$-modes \citep{Seljak1997, 1997PhRvD..55.1830Z, 1997PhRvD..55.7368K}.
Precision measurements of the angular power spectrum of the intensity fluctuations, \cltt, are a cornerstone of our current understanding of cosmology.
The power spectrum of the primordial $E$-modes, \clee, has also been well-characterized, as has their relationship with the temperature anisotropy pattern, \clte. 
These measurements are consistent with a single source for both the temperature and $E$-mode signals -- the adiabatic density fluctuations in the primordial universe \citep{2014ApJ...783...67B,2005ApJ...624...10L,2006ApJ...647..813M, brown2009, quiet_FirstSeason, quiet_SecondSeason, 2013ApJS..208...20B}.

Unlike $E$-modes, odd-parity $B$-mode patterns are not produced by scalar fluctuations.
Any primordial $B$-modes would be evidence for tensor or vector perturbations in the gravitational metric when the CMB was emitted \citep{1997PhRvL..78.2054S, 1997PhRvL..78.2058K}.
The only source of such perturbations in the standard cosmological model is a period of cosmic inflation in the early universe. 
If cosmic inflation occurred, the induced tensor perturbations (or inflationary gravitational wave background) would imprint a Gaussian field of $B$-mode polarization on the primordial CMB that is largest at degree angular scales.
The angular power spectrum of those primordial $B$-modes, \clbb, would then provide information about cosmic inflation, and could be a window into physics at grand-unified energy scales, when the electroweak and strong forces are expected to unify \citep{KamionkowskiKosowski}.

The CMB radiation that we observe has been modified by secondary effects as compared with the primordial signal.  
One such effect is the gravitational lensing of CMB photons by cosmological large-scale structure (LSS). 
The gravitational structure traversed by the CMB radiation as it travels to us distorts the primordial CMB temperature and polarization fluctuations, and generates $B$-mode polarization anisotropies \citep{HuOkamoto2002, 2006PhR...429....1L}. 
This distortion adds secondary power to the $B$-mode angular power spectrum peaking at scales of $0.2^\circ$, and also imprints a non-Gaussian correlation between anisotropies in the CMB temperature and polarization. 
That correlation can be used to reconstruct maps of the integrated structure along the line of sight -- all the structure in the observable universe. 
High fidelity maps of this effect will be a powerful probe of fundamental physics, cosmology, and extragalactic astrophysics. 
These maps will also enable the removal of this secondary $B$-mode signal to precisely characterize any primordial \clbb\ due to inflationary gravitational waves \citep{seljak2004}.

The scientific prospects from precise characterization of LSS using CMB lensing are significant. 
The neutrino mass, known to be non-zero from neutrino flavor oscillation measurements, can be measured by its effect on LSS formation.
The high velocities of cosmic neutrinos inhibit gravitational clustering on small scales, suppressing LSS on scales smaller than $\sim 100$\,Mpc.
Measurements of the gravitationally lensed CMB in both temperature and polarization have the potential to measure the sum of neutrino masses with an uncertainty comparable to the known mass splittings measured by flavor oscillation experiments, which set the minimum sum of the neutrino masses to be 58\,meV \citep{beringer2012}.  
Also, CMB lensing measurements break the degeneracy that exists in primordial CMB temperature anisotropies between curvature, dark energy parameters, and the sum of neutrino masses \citep{SmithHuKaplinghat_2006,calabrese2009,smith09}, and thus improve constraints on those parameters.
In particular, CMB lensing measurements are complementary to other probes of dark energy because they are sensitive to its high-redshift behavior \citep{calabrese2011}.
Cross-correlating LSS maps from CMB lensing measurements with other LSS mass-tracing probes will improve the calibration of these tracers and the statistical accuracy of the resulting cosmological constraints.

Lensing of the CMB by LSS was first observed in temperature measurements, through the LSS-induced non-Gaussianities \citep{smith2007,das2011,das2013_arXiv,vanEngelen2012,planck2013_XVII_arXiv}, and modification of the power spectrum \cltt\ \citep{calabrese2008,reichardt2009,das2013_arXiv,story2012_arXiv,planck_XVI_arXiv}.
Polarization measurements have the potential to more precisely reconstruct the LSS-induced signal because the lensing $B$-modes are not contaminated by large primordial CMB fluctuations \citep{HuOkamoto2002}.
Measurement of non-Gaussianity in the CMB polarization was first reported recently by \sptpol\ and \pb\ using cross-correlation with Herschel observations of high-redshift galaxies \citep{hanson2013,PB_GalaxyCross_2014}. 
While the cross-correlation studies establish the existence of $B$-modes from gravitational lensing, they are not sensitive to LSS throughout cosmic time because they rely on tracers that exist over a limited range of redshift. 
Recently \pb\ reported a measurement of the non-Gaussianities induced by LSS in the polarized CMB from CMB data alone, which is therefore sensitive to all lensing distortions along the line of sight \citep{PB_CLdd_2014}.

These early measurements imply an amplitude of \clbb, but a detection of \clbb\ \textit{itself} has not yet been published. 
The small amplitude of this signal compared to other sources of anisotropy in the CMB makes it very difficult to measure without contamination from the instrument or astrophysical foregrounds.
In this sense, \clbb\ is more difficult to characterize than the non-Gaussianity induced by LSS, but its precise characterization is required to search for the signal from cosmic inflation, and extract all of the science possible from $B$-mode cosmology.
In this paper we present a measurement of \clbb\ using \pb. 

The \pb\ experiment uses a millimeter-wave polarimeter to make deep maps of the CMB temperature and polarization anisotropies. 
\Secref~\ref{sec:instrument} describes the instrument, and \secref~\ref{sec:obs} details the observations that were performed to obtain the data reported here. 
\Secref~\ref{sec:calibration} describes the calibration of these data, and \secref~\ref{sec:analysis} describes the analyses we used to produce the \clbb\ measurement. 
Possible sources of systematic contamination of \clbb\  are evaluated in \secref~\ref{sec:fg} -- astrophysical foregrounds -- and \secref~\ref{sec:systematics} -- instrumental systematics, and found to be small compared to the measured signal. 
Finally, we present the \pb\ measurement of binned \clbb\ power over angular multipoles $500 < \ell < 2100$ in \secref~\ref{sec:results}, and conclude with a discussion of the measurement in \secref~\ref{sec:summary}.

%%%%%%%%%%%%%%%%%%%%
\section{Instrument}
\label{sec:instrument}

The \pb\ experiment is composed of a two-mirror reflective telescope -- the Huan Tran Telescope (HTT) -- coupled to a cryogenic receiver. This instrument is installed at the James Ax Observatory in the Atacama Desert in Northern Chile.  

\subsection{Telescope}
\label{sec:inst-telescope}

The HTT reflectors are in an off-axis Gregorian configuration satisfying the Mizuguchi-Dragone condition. 
This design provides low cross-polarization and astigmatism over the diffraction-limited field of view \citep{mizuguchi,dragone,Tran:2008p1693}. 
The telescope optics, cryogenic receiver, and electronics are installed on a mount that provides control of telescope pointing in azimuth and elevation.
The primary mirror is an off-axis paraboloid comprised of a central panel with 50 micron RMS surface accuracy, and eight lower-precision outer panel segments.
Projected along boresight, the aperture is an ellipse with a 3.5\,m minor axis (2.5\,m for the central panel). 
A 4\,K aperture stop in the receiver creates a 2.5\,m primary illumination pattern on only the central monolithic panel, giving a beam size of 3.5\arcmin\ full-width at half-maximum (FWHM). 
The outer panels ensure that any spillover illumination is filled by radiation from the comparably faint sky, rather than radiation from the telescope structure or ground. 
Several co-moving baffles serve to reduce side lobe response.
Additional absorptive shielding above the primary mirror (extending about 1\,m in radius) was installed in January 2013, eliminating a single localized far sidelobe with an intensity of about -50\,dB. 
\Secref~\ref{sec:sys-ground} describes the scan synchronous signal filtering that removes contamination from signals fixed in azimuth; we find that this filtering effectively removes the signal created by this sidelobe observing the ground. 
\secref~\ref{sec:sys-null} describes null tests used to demonstrate that the $B$-mode spectra measured before and after the additional shielding installation are consistent.

\subsection{Receiver}

The cryogenic receiver coupled to the HTT houses a cold half-wave plate~(HWP), re-imaging optics, aperture stop, and a focal plane of \npix\ dual-polarization pixels (\nbolo\ detectors) with a 2.4\textdegree\ diameter field of view \citep{Kermish_SPIE2012}. 
The detector array is composed of seven individual wafers, each of which is a hexagon 80\,mm across \citep{Arnold_SPIE2012}.
The cold stop and warm co-moving telescope baffles shield the detectors from stray radiation. 
Radiation incident on a pixel is coupled from free space to superconducting microstrip wave guides by a dual-polarization antenna and contacting dielectric lenslet.
The spectral bandpass of each detector in a pixel is determined by three-pole microstrip filters; the design band is centered at 148 GHz with 26\% fractional integrated bandwidth.
The power transmitted through the filter is deposited on a superconducting transition-edge sensor~(TES) bolometer.
Each TES bolometer is biased with an AC voltage to keep it at the superconducting temperature; a change in the optical power on a bolometer creates a compensating change in electrical current, which is measured using frequency-domain multiplexed superconducting quantum interference device (SQUID) ammeters \citep{Dobbs2012}.

\subsection{Site}

The HTT is located at the James Ax Observatory in Northern Chile on Cerro Toco at West longitude 67\textdegree 47\arcmin 10.40\arcsec, South latitude 22\textdegree 57\arcmin 29.03\arcsec, elevation 5,200\,m. 
The median precipitable water vapor~(PWV) during the first season of \pb\ observations, when the nearby \apex\ water vapor radiometer was operating, was 1.0\,mm. 
This corresponds to a sky brightness in the \pb\ design band of 12\,K at an elevation angle of 60\textdegree.

%%%%%%%%%%%%%%%%%%%%
\section{Observations}
\label{sec:obs}

\Pb\ observations began in January 2012. After validation and initial calibration of the instrument, regular scientific observations of the CMB began in June 2012; the data reported here were collected between that time and June 2013. 
During this period, 2,400 hours of calibrated CMB observations and 400 hours of calibration data were recorded. 
In this section, we describe the \pb\ observation strategy. 
In \secref~\ref{sec:cuts}, we describe the instrument and weather-related data selection criteria that define the data set used for this analysis.
Averaged across these observations, the array sensitivity in the science band (see \secref~\ref{sec:mapmaking}) is $23$\,\microKrtsec.

\subsection{Patches}

\begin{figure}[htpb]
\begin{center}
\includegraphics[width=3in]{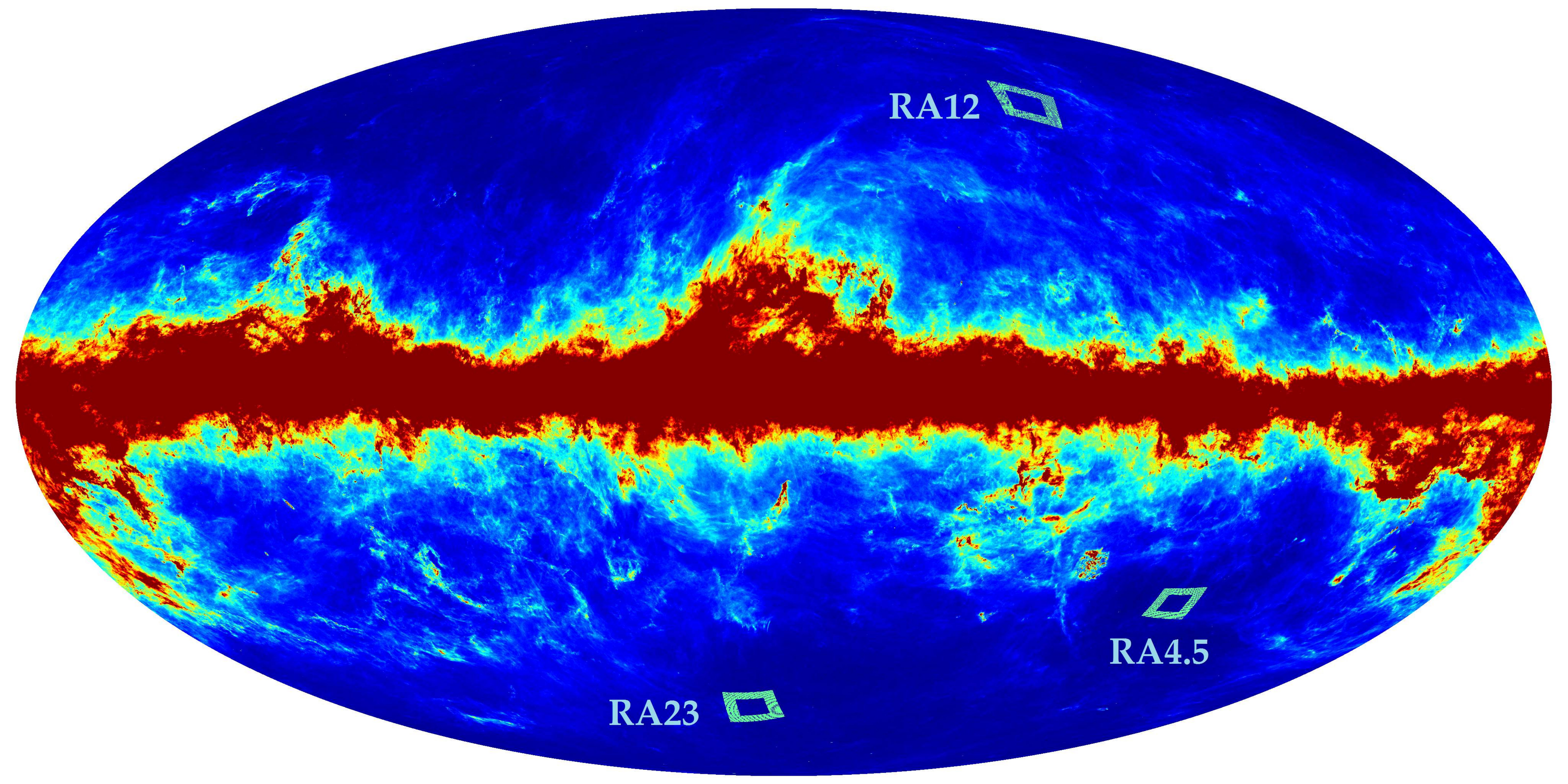}
\end{center}
\caption{\label{fig:patches}The three \pb\ patches overlaid on a full-sky 857\,GHz intensity map \citep{2013arXiv1303.5062P}. Patches were chosen for low dust emission, overlap with other observations, and to allow nearly continuous CMB observations from the James Ax Observatory in Chile.}
\end{figure}

The results reported here are from observations of three patches of sky.
For each \pb\ patch, the patch center coordinates and effective areas (as defined in \secref~\ref{sec:power_spec_error_bars}) are given in \tabref~\ref{tab:patches}. 
The patches are referred to here as RA4.5, RA12, RA23, following their right ascension.
The locations of the patches are shown in Figure \ref{fig:patches} overplotted on a full-sky 857\,GHz intensity map tracing the signal from galactic dust \citep{2013arXiv1303.5062P}.
The patch locations are chosen to optimize a combination of low dust intensity, availability throughout the day, and overlap with other observations for cross-correlation studies.
RA23 and RA12 were selected to overlap with Herschel-ATLAS observations, and RA4.5 and RA23 to overlap with \quiet\ observations \citep{Eales_2010,quiet_SecondSeason}.

\begin{table}[htpb]
\begin{center}
\caption{\label{tab:patches}The three \pb\ patches.}
\begin{tabular}{lccc}
\hline
\hline
Patch & RA  & Dec & Effective Area \\
\hline
RA4.5 & 4\rahrs 40\ramin 12\rasec & -45\textdegree \phantom{00\arcmin} & 7.0 \sqdeg \\
RA12 & 11\rahrs 53\ramin 0\rasec & -0\textdegree 30\arcmin & 8.7 \sqdeg \\
RA23 & 23\rahrs 1\ramin 48\rasec & -32\textdegree 48\arcmin & 8.8 \sqdeg
\end{tabular}
\end{center}
\end{table}

\subsection{Scan pattern}

\pb\ observes a patch continuously for up to eight hours while that patch is above the minimum observing elevation.  
For the first three months of observations, that observing horizon was set at 40\textdegree\ elevation.  
For the remainder of the data, the observing horizon was 30\textdegree.  
The patches rise to a maximum elevation angle of 80\textdegree.  
To optimize sensitivity and linearity in changing atmospheric loading, the detectors are re-biased every hour.  
Before and after each re-bias we measure the relative gain of the detectors using both elevation nods (2\textdegree\ modulations of the telescope elevation angle) and observation of a chopped 700\textdegree C thermal source visible through a small hole in the secondary mirror.
The chopping frequency for the thermal source calibration is stepped between 4 and 44 Hz to simultaneously measure the detector time constant.

Observation of one patch is broken into 15-minute scans at constant elevation, during which the telescope scans back and forth in azimuth $3^\circ$
at a speed of $0.75^\circ/{\rm s}$ on the sky.
The telescope then moves in azimuth and elevation to where the patch will be in $7.5$ minutes, and the constant elevation scan (CES) pattern is repeated.

In one CES there are approximately 200 constant-velocity subscans; data between subscans when the telescope is accelerating, 36\% of the CES, are discarded for this analysis.
Observing at constant elevation allows scan-synchronous systematic signals, such as ground pick-up, to be removed from the map with only a small loss of information.

The cold HWP is always stationary during observations. 
Over the first half of the season, the HWP was stepped in angle by 11.25\textdegree\ every 1--2 days. 
During the second half of the season the HWP was stepped in angle only occasionally, as we worked to characterize the HWP-dependent signals in the data. 
As described in \secref~\ref{sec:polmodel}, the HWP was important in understanding the polarization angles of the detectors, and provided the ability to constrain the pixel-pair relative gain (see \secref~\ref{sec:sys-special-diffgainmap}).

\subsection{Yield}
\label{sec:obs_yield}
Of the \nbolo\ optical TES bolometers in the \Pb\ focal plane, 1,015 were able to be electrically biased and showed nominal optical response to astrophysical point sources. 
During observations, readout channels that show anomalously high noise properties are turned off so that pathological noise effects are not induced in other detectors. 
Individual pixels are permanently excluded when they show no optical response in either one or two of the bolometers, high differential pointing, high differential gain, high variation in differential gain, or a large uncertainty on the polarization angle calibration.
This leaves 373 pixels (746 bolometers) that are used in the reported measurement. 
Further data selection criteria are described in \secref~\ref{sec:cuts}.

%%%%%%%%%%%%%%%%%%%%
\section{Calibration}
\label{sec:calibration}

As input to map-making and power spectrum estimation, there are four primary instrument properties to be modeled: individual detector pointing, thermal-response calibration, polarization angle, and the instrument effective beam. 
The models for these properties are described in the following section. 
Uncertainties in these models are evaluated in \secref~\ref{sec:systematics}, and none are found to produce significant contaminant signals with respect to the detected \clbb\ signal.

\subsection{Pointing}
\label{sec:pointing}

A five-parameter pointing model \citep{magnum2001} characterizes the relationship between the telescope's encoder readings and its true boresight pointing on the sky. Of the parameters described in this reference, \pb\ uses IA, the azimuth encoder zero offset, IE, the elevation encoder zero offset, CA, the collimation error of the electromagnetic axis, AN, the azimuth axis offset/misalignment (north-south) and AW, the azimuth offset/misalignment (east-west). 
We experimented with extending and modifying this parameter set, and did not find substantial improvements to the model.  
The pointing model is created by observing bright extended and point-like millimeter sources that were selected from known source catalogs \citep{vlanorth, at20g} to span a wide range in azimuth and elevation.
These pointing observations occurred several times per week during observations.
The best-fit pointing model recovers the source positions for the sources that were used to create it with an accuracy of 25\arcsec\ RMS.

Individual detector beam offsets are determined relative to the boresight using raster scans across Saturn and Jupiter. 
These beam offsets are then combined with the boresight pointing model to determine the absolute pointing of each bolometer. 
The offsets were typically measured several times per week during observations throughout the season, showing an RMS fluctuation of less than 6\arcsec\ over time. 
The offsets show arcsecond-level differential pointing between the two detectors in a pixel, which is shown to be a negligible contaminant in \secref~\ref{sec:systematics}.

The robustness of the pointing model is tested by fitting the same model to various subsets of the pointing data separated by source, time, or environmental conditions. 
Some systematic differences were observed, and we believe this indicates a problem with the pointing model that is not well-characterized by the residual source position accuracy.
As described in \secref~\ref{sec:sys-abspoint}, these systematic pointing uncertainties were simulated and they were found to not create significant bias in \clbb. 
However, the systematic uncertainties in our pointing model lead to the reduction of our sensitivity to high-$\ell$ anisotropies by increasing the effective beam width, blurring the maps slightly. 
This blurring could be different for each of the patches, because the inaccuracies in the pointing model are a function of telescope azimuth and elevation angle, and each patch traverses a different path in azimuth and elevation.
This blurring is evaluated by measuring the observed width of radio-bright point sources in the co-added CMB temperature maps, as discussed in \secref~\ref{sec:pdbeams}, where we determine that this blurring is not important in the characterization of \clbb\ reported here.

\subsection{Beam}
\label{sec:beammodel}

The basis for the beam model is measured by computing the angular response spectra of several maps of Jupiter created using the planet observations described in \secref~\ref{sec:pointing}. 
It is then modified to account for blurring in the maps due to pointing model inaccuracies, evaluated using point sources in the patches.

\subsubsection{Jupiter measurements}

Individual detector maps are made from timestreams that are filtered to reduce the effect of atmospheric fluctuations. 
This filtering is accomplished with a masked polynomial: a first-order polynomial that is fit to the timestream everywhere outside of a 50\arcmin\ radius around the planet.  
The individual detector maps are combined to create a single-observation map with adequate coverage. 
The weighting used to combine individual detector maps is a noise weighting calculated from the RMS of the map outside of the mask radius.  

The azimuthally-averaged Fourier components are then found for each single-observation map.  
We work in the flat-sky approximation, appropriate for our small patch sizes (each patch represents less than 0.03\% of the sky). 
The two-dimensional Fourier transform of the temperature-calibrated map, $M(\ell,\phi_{\ell})$, is averaged over $\phi_\ell$

\begin{equation}
\tilde{M_{\ell}} = \mathbf{Re}\Bigg\{ \frac{1}{2\pi}\int M(\ell, \phi_{\ell})d\phi_{\ell}\Bigg\},
\end{equation}

\noindent and binned in $\ell$, creating bins of width $\Delta\ell\approx80$. 
The beam can be approximated as azimuthally symmetric in intensity because of the rotation of the instrument beams as projected on the CMB patch due to changes in patch parallactic angle over the course of the day from our mid-latitude site \citep{Fosalba_2002}.

The underlying beam model for each observation is found by dividing the azimuthally-averaged map transforms by both the angular Fourier response function for a finite planetary disk, $T_\ell$, and a correction for the map pixel size used, $W_\ell$.  
The pixel window function for a pixel of size $\Delta$ is given by \citet{Wu_2001} as

\begin{equation}
W_{\ell} = e^{-(\ell\Delta)^{2.04}/18.1} \left[ 1-(2.72\times 10^{-2})({\ell}\Delta)^2 \right ],
\end{equation}
and $T_\ell$ of a planetary disk of known radius $R$ and temperature $T_p$ is 
\begin{equation}
T_{\ell} = 2\pi T_p R^2 J_1(\ell R)/(\ell R),
\end{equation}
where $J_1$ is the Bessel function of the first kind of order unity.

The multipole components and covariance matrix of the beam are estimated by the inverse-noise-weighted mean values and covariance matrix of the power spectra of many planet observations. 
The resulting $B_{\ell}$ and its uncertainty in each multipole bin from observations of Jupiter are shown in \figref~\ref{fig:eff_bl}.  
Results using Saturn observations are consistent.

\subsubsection{Point sources in the CMB patches}
\label{sec:pdbeams}

\begin{figure}
\begin{center}
\includegraphics[width=3.5in]{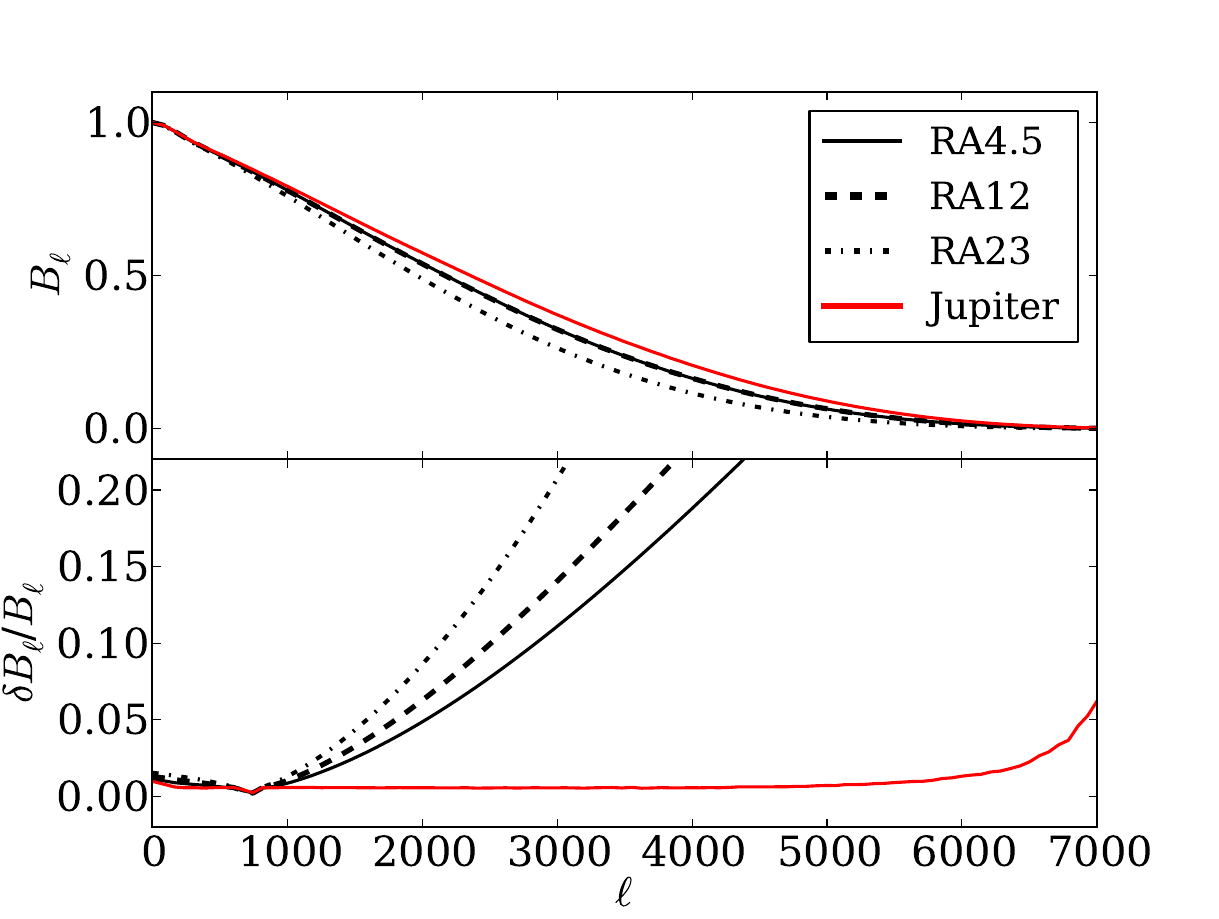}
\caption{Top: Beam profiles measured from Jupiter (red) and from fitting point sources with a Gaussian-smoothed Jupiter beam (black). Bottom: Beam uncertainties given as 1-$\sigma$ uncertainties normalized by the beam profile.}
\label{fig:eff_bl}
\end{center}
\end{figure}

The effective beam $B_{\ell}^\mathrm{eff}$ used in our power spectrum analysis accounts for blurring of the co-added CMB maps that is introduced by inaccuracies in our pointing model.  
We model the effect of an RMS pointing error in map space, $\sigma_p$, as 

\begin{equation}
B_{\ell}^{\mathrm{eff}} = B_{\ell} \times\ e^{-{\ell}({\ell}+1)\sigma_p^2/2},
\end{equation}
where  $\sigma_p$ for each field is estimated by fitting each point source with the Jupiter beam profile convolved with a Gaussian smoothing kernel.  
We estimate a $\sigma_p$ likelihood function for each individual point source in each map.
We then combine the sources within each patch to find a joint likelihood for the RMS pointing error on that patch. 
The results are shown in \tabref~\ref{tb:ps_pointing} and the patch-specific beam functions are displayed in \figref~\ref{fig:eff_bl}. 
Using these beams for each patch, and including the beam covariance, the individual patch \cltt\ power spectra, and the patch-combined power spectra (shown in \figref~\ref{fig:spectrumresults}) are consistent with \wmap-9 \lcdm\ \citep{2013ApJS..208...20B}, as described in \secref~\ref{sec:self_calibration}. 

We increased the uncertainty in $B_\ell^{\rm eff}$ to account for discrepancies in the measurements of $\sigma_p$.
The measurements of $\sigma_p$ from the ten point sources in RA4.5 exhibited statistically significant differences with one another, implying that the amount of blurring is not constant throughout the map, or that the model for the blurring is not capturing the entire effect; we do not understand the origin of this effect. 

The beam uncertainty increases the uncertainty of our absolute gain calibration to \cltt\ (see \secref~\ref{sec:gain}) by a factor of 1.5, from 2.8\% to 4.1\%. 
The differences and uncertainties in the patch-specific beam functions primarily affect the $\ell$-range at the high end of our reported band powers and beyond, having little effect on the constraint of \ABB\ given in \secref~\ref{sec:results}, where most of the significance comes from the low end of the reported $\ell$-range. 
An analysis using simply the Jupiter beams, calibrated using \cltt, results in an \ABB\ that differs by 0.3\% from the reported result.

\begin{table}
\begin{center}
\caption{Pointing error for \Pb\ observing fields}
\label{tb:ps_pointing}
\begin{tabular}{ l c }
\hline
\hline
Field & RMS pointing error [\arcsec] \\ 
\hline
RA23 & 64.5 $\pm$ 20.9 \\
RA12 &  26.7 $\pm$ 18.2 \\
RA4.5 & 31.5 $\pm$ 15.3 \\
\hline
\end{tabular}
\end{center}
\end{table}

\subsection{Gain}
\label{sec:gain}

Bolometer Time-Ordered Data (TOD) represent electrical current in the detector.
A measurement of the responsivity of the detector, or gain, is required to convert current into CMB temperature units. 
An array of astrophysical and ground-based calibrators are used to measure the gain. 
The CMB \cltt\ power spectrum is used to determine a single scale factor that provides an absolute gain calibration for the instrument; all other calibration sources are used as relative calibrators. 

A small chopped thermal source is the primary relative calibrator. 
This source couples into the detector beam through a 6mm light-pipe penetrating the secondary mirror.
The thermal source has a regulated temperature of 700\textdegree C and uses a chopper wheel to modulate at a series of six frequencies between 4\,Hz and 44\,Hz during each calibration. 
Calibrations occur at the beginning and end of each hour-long CMB observation block. 
Bolometer voltage biases are fixed during these hour-long blocks. 
The frequency response of each bolometer to this calibrator is well fit by a single heat capacity model with a time constant of 1--3\,ms. 

The polarized response of individual detectors to the thermal source is characterized by referencing to measurements of an astrophysical source at different HWP rotation angles. 
Characterizing the polarization of the thermal source is particularly important because it directly impacts the relative gain of the two bolometers in a pixel pair in a HWP rotation angle dependent way.
A miscalibration of relative gain can result in systematic leakage of CMB temperature to polarization.  In \secrefs~ \ref{sec:sys-relgainsim} and \ref{sec:sys-special} we test our relative gain model and show that the systematic uncertainties associated with it are small.

\subsection{Polarization}
\label{sec:polmodel}

Each of the \pb\ detectors' response to polarized signal, including its polarization angle, polarization efficiency, and leakage due to relative-gain miscalibration, is modeled. 
Also modeled are non-idealities in the HWP, which may vary for different detectors across the focal plane.
This section describes the polarization model as developed using two astrophysical calibrators.  Taurus A (\taua) is a polarized supernova remnant that was first used for polarization angle calibration by COMPASS \citep{farese2004} and was later characterized by \wmap\ between 23 and 94\,GHz \citep{2011ApJS..192...19W}, and by the IRAM 30\,m telescope at 90\,GHz \citep{2010A&A...514A..70A}.
Centaurus A (\cena) is a fainter polarized radio-bright galaxy that has been characterized by \quadexp\ \citep{2010ApJ...710.1541Z}; it is used as a consistency check.
The polarization angle of each detector relative to the instrumental reference frame is found from this model.  Section \ref{sec:self_calibration} describes calibration of the overall instrument polarization angle using the CMB itself.

\subsubsection{Polarization model characterization using Tau A}

\begin{figure}
 \begin{center}    \includegraphics[width=3.5in]{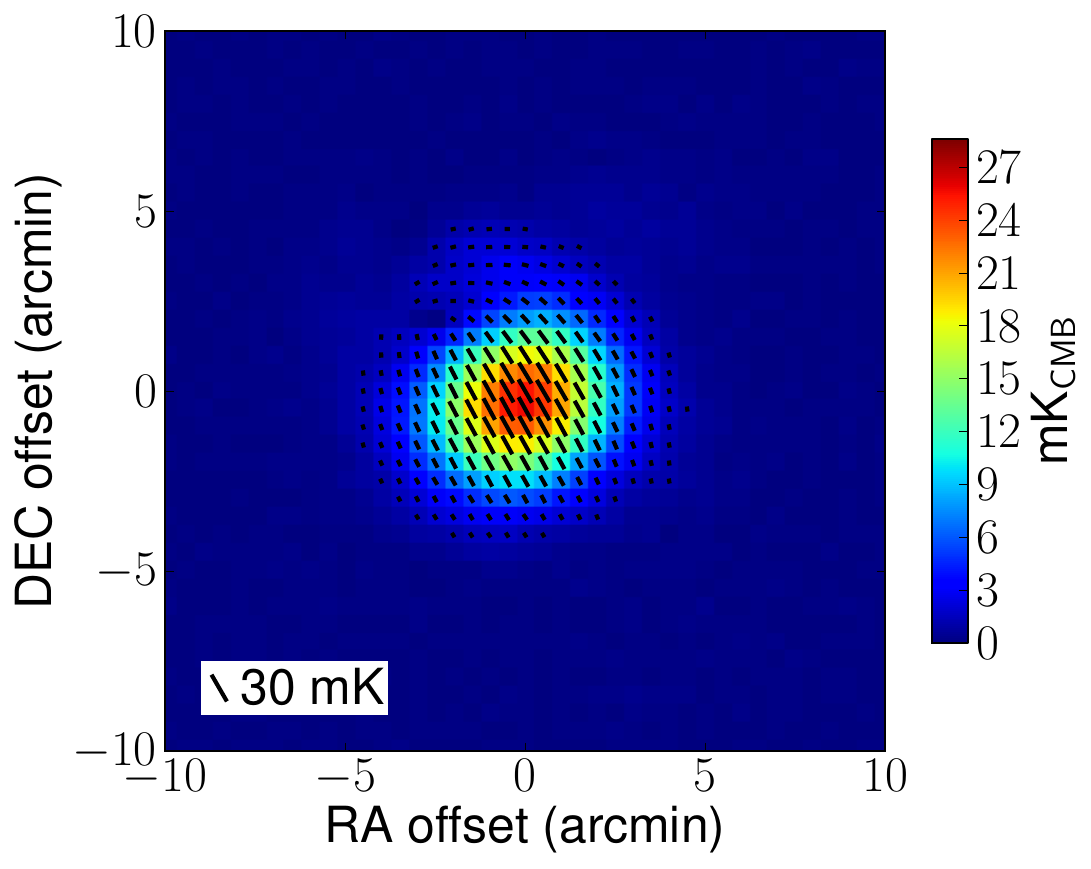}
  \caption{Full season co-added polarization intensity map of Tau A observed by \pb. The orientations of bars in map pixels represent polarization angles at each map pixel.}
  \label{fig:coaddedmap_taua}
 \end{center}
\end{figure}

\pb\ observed \taua\ several times per week throughout the observations reported here, except when \taua\ was within 30\textdegree\ of the sun, which occurred between May and July of each year. 
This resulted in 125 \taua\ observations which were evaluated based on the same selection criteria as were implemented in the CMB analysis, where applicable (see \secref~\ref{sec:cuts}).
The observation pattern is a raster scan where the telescope tracks \taua\ while executing a set of constant-velocity subscans with steps in elevation between them.
Figure \ref{fig:coaddedmap_taua} shows a full-season co-added polarization map of \taua.

\taua\ is an extended source; IRAM observed that \taua's polarization angle and polarized intensity are not uniform over its spatial extent.
To compare the IRAM maps to \pb\ observations, a simulated timestream is generated for each bolometer by scanning the IRAM map with that bolometer's pointing. 
To create each sample of simulated timestream, the IRAM maps are convolved with that detector's beam, modeled for this as an elliptical Gaussian (Note that the IRAM 30\arcsec\ beam is negligibly small compared to the \pb\ beam). 
The simulated timestreams $s_{i,\parallel\perp}(t)$ are calculated from the IRAM $I$, $Q$, and $U$ maps:

\begin{eqnarray}
s_{i,\pp} (t) & = &  I_{{\rm sim},i,\pp}(t) + Q_{{\rm sim},i,\pp} (t) \cos 2\Theta_{i,\pp(t)} \nonumber \\
   & &  + U_{{\rm sim},i,\pp} (t) \sin 2\Theta_{i,\pp}(t),
\end{eqnarray}

\noindent where $\Theta_{i,\parallel\perp}(t)$ is the polarization angle
projected on the sky in equatorial coordinates for the two orthogonal
detectors ($\parallel\perp$) of pixel $i$. 
This is related to the detector polarization angle $\theta_{i,\parallel\perp}$, the HWP angle $\theta_{\rm HWP}$, and parallactic angle $\theta_{\rm PA}$ by

\begin{equation}
 \Theta_{i,\pp}(t) = \left(\frac{\pi}{2} - \theta_{i,\pp} \right) + 2\theta_{\rm HWP}(t)+ \theta_{\rm PA}(t), \label{total_polarization_angle}
\end{equation}

\noindent and the simulated intensity and polarization signals $X_{\rm sim}(t) \in { ( I_{\rm sim}(t),Q_{\rm sim}(t),U_{\rm sim}(t) ) }$, which are calculated from the IRAM maps $X(\boldsymbol{p})$ and the detector beam $B_{i,\parallel\perp}$: 

\begin{eqnarray}
X_{{\rm sim}, i,\pp} (t) & = & \int X(\boldsymbol{p}) B_{i,\pp}(\boldsymbol{p} - \boldsymbol{p}_{i,\pp}(t)) d\boldsymbol{p}, \nonumber \\
   & & X\in{I,Q,U}.
\end{eqnarray}

Polarization timestreams are created from \pb\ data by differencing the two orthogonally oriented detectors within one pixel (see \secref~\ref{sec:mapmaking} for the CMB polarization map-making process). 
Simulated difference timestreams of \taua\ observations are calculated from individual detector timestreams $s_{i,\parallel\perp}$ using the fact that $\Theta_{i,\perp}=\Theta_{i,\parallel}+\pi/2$.

\begin{eqnarray}
 \label{eq:model_diff}
  d_{i} (t) &=& \frac{1}{2}(s_{i,\parallel}-s_{i,\perp}) \nonumber \\
 & \approx &
 \frac{1}{2}\left[I_{{\rm sim},i,\parallel}(t) - I_{{\rm sim},i,\perp}(t)\right] \nonumber \\ 
  &&+ \epsilon_i \left[Q_{{\rm sim},i}(t)\cos2\Theta_i(t) +
   U_{{\rm sim},i}(t)\sin2\Theta_i(t)\right] \nonumber \\
 &&+ \frac{\Delta g_i}{2} \left(I_{{\rm sim},i,\parallel}(t) 
 + I_{{\rm sim},i,\perp}(t)\right).
\end{eqnarray}

\noindent Here we have introduced the polarization efficiency $\epsilon_i$ and pixel-pair relative-gain error $\Delta g_i$.

The constant-velocity portion of each subscan is fit to a masked polynomial, as with observations of planets.
In this case, the polynomial is fifth-order, and the mask is 10\arcmin\ in radius from the center of \taua. 
The best-fit polynomial is then subtracted from the data. Detector differencing and polynomial baseline subtraction both act to remove atmospheric signals. The simulated observations of the IRAM maps undergo the same filtering, and then the data are fit to the simulations for detector pointing within 5\arcmin\ of the center of \taua. The fit parameters for each pixel are the pixel polarization angle $\theta_i$ (implicitly in $\Theta_i$), polarization efficiency $\epsilon_i$, and the pixel-pair relative gain error $\Delta g_i$.

The polarization efficiency predicted due to the expected performance of the HWP over the finite \pb\ spectral band is between 0.97 and 0.98, slightly different from wafer to wafer due to the different measured spectral bands.
The mean pixel polarization efficiency from the \taua\ fit is $0.994 \pm 0.036$. 
Note that this measured polarization efficiency is strongly affected by any difference in polarization fraction between 90 and 148\,GHz. 
In CMB map-making, the mean theoretical value of 0.976 is used as the polarization efficiency for every detector. The systematic uncertainty is constrained by the difference between the prediction and the mean value of the \taua\ fit, which then encompasses other possible uncertainties than the HWP.

To monitor the time stability of the detector polarization angle and the accuracy of the reported HWP angle, the detector polarization angles fit to the full season of \taua\ data, as described above, are used to make maps of Tau A for each observation. 
In this  map-making process, pixel-sum and pixel-difference subscans are filtered with the same polynomial baseline subtraction as in the pixel polarization angle fit. 
Data from every pixel are co-added, and the polarization angle of Tau A for that observation, $\alpha_{\mbox{\scriptsize{Tau A}}}$, is calculated from that observation's co-added $Q$ and $U$ maps:

\begin{equation}
  \label{eq:daily_polangle}
 \alpha_{\mbox{\scriptsize{Tau A}}} = \frac{1}{2}\arctan{\left(\frac{\sum{U_j}}{\sum{Q_j}}\right)},
\end{equation}

\noindent where $j$ is a map pixel within 10\arcmin\ of the center of \taua.
The variation of this polarization angle over the season is 1.2$^\circ$ (RMS). 

The polarization model is based on a hierarchical understanding of our polarization calibration, consisting of a global reference polarization angle, the wafer-averaged polarization angles relative to that global angle, and the individual pixel angles relative to the wafer-averaged angle. We consider uncertainty in each of these. 
The systematic uncertainty in the global reference angle and wafer-averaged angle are shown in \tabref~\ref{tb:syserr_taua}. These are dominated by uncertainties in the pixel-pair relative gain and in the non-axisymmetric beam model and the substructure of Tau A at 148 GHz. Non-idealities in the HWP over the finite \pb\ spectral bandwidth are also an important source of uncertainty, both in rotation angle of linear polarization and in the mixing of circular polarization into linear polarization.
Using the upper limit on \taua's  circular polarization fraction of 0.2\% \citep{2011A&A...528A..11W}, the systematic error from the circular polarization of \taua\ is estimated to be 0.09\textdegree.
The individual pixel polarization angle uncertainty in each wafer is estimated to be 1.0\textdegree\ from the spread of the pixel polarization angle distribution from the \taua\ measurement.
The other systematic effects we evaluated, listed in \tabref~\ref{tb:syserr_taua}, are all negligible. 
The impact of all these uncertainties on the \clbb\ and \cleb\ power spectra are addressed in \secref~\ref{sec:systematics}.

\begin{table}
\begin{center}
\caption{Systematic uncertainties in global reference and wafer-averaged polarization angle, as measured using \taua. 
Uncertainty in the global reference angle as measured using \cleb\ is addressed in \secref~\ref{sec:self_calibration}.}
\label{tb:syserr_taua}
\begin{tabular}{ l c c}
\hline
\hline
  Angle Uncertainty & Global Reference & Wafer-Averaged \\ 
\hline
Absolute pointing uncertainties & 0.12\textdegree & - \\
Beam uncertainties & 0.21\textdegree & 0.23\textdegree\\
Relative gain uncertainties & 0.22\textdegree & 0.42\textdegree \\
Non-ideality of HWP & 0.21\textdegree & 0.64\textdegree \\
Circular polarization of Tau A & 0.09\textdegree & $\ll$ 0.1\textdegree \\
HWP angle uncertainties & 0.15\textdegree & 0.13\textdegree\\
Pixel pointing uncertainties & $\ll$ 0.1\textdegree & 0.18\textdegree\\
Bolometer time constant & $\ll$ 0.1\textdegree & $\ll$ 0.1\textdegree\\
Filtering effect & $\ll$ 0.1\textdegree & $\ll$ 0.1\textdegree\\
Polarized dust & $\ll$ 0.1\textdegree & $\ll$ 0.1\textdegree\\
\hline
Total & 0.43\textdegree & 0.83\textdegree\\
\end{tabular}
\end{center}
\end{table}

\subsubsection{Consistency check with Cen A}

\pb\ observations of \cena\ follow the same raster scan as the observations of \taua. 
Single-observation maps of \cena\ are produced and the polarization angle of \cena\ is measured from those maps as was done for \taua\ using Equation~\ref{eq:daily_polangle}. To calculate the polarization angle of \cena, all map pixels within 12\arcmin\ of its center are used. 
The \quadexp\ experiment measured the polarization of \cena\ at 100 and 150\,GHz \citep{2010ApJ...710.1541Z}. 
\pb\ measured a \cena\ polarization angle of $147.9^\circ\ \pm 0.6^\circ ({\rm stat.}) \pm 1.0^\circ ({\rm sys.})$ 
with the \taua-derived polarization angle.
Using the \cleb-derived polarization angle described in \secref~\ref{sec:self_calibration} 
results in a measured \cena\ polarization angle of $149.0^\circ\ \pm 0.6^\circ ({\rm stat.}) \pm 0.9^\circ ({\rm sys.})$. 
The \pb\ and \quadexp\ measurements of \cena\ agree within their measurement uncertainties.

\subsection{Calibration using the CMB}
\label{sec:self_calibration}

\begin{figure}[htpb]
 \centering
 \includegraphics[width=3.4in]{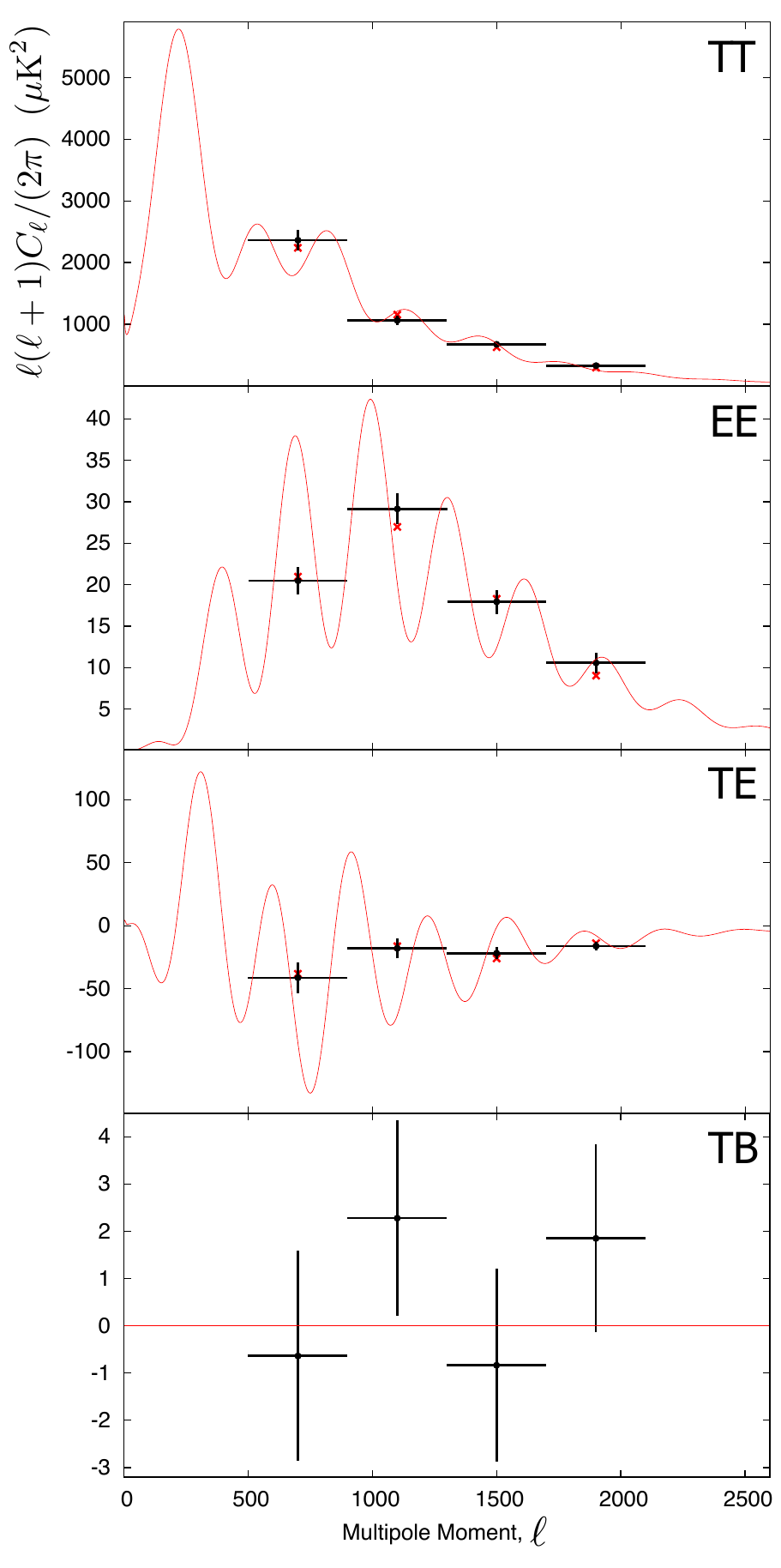}
 \caption{\label{fig:spectrumresults}First season \pb\ power spectra used for calibration and cross-checks of the calibration. 
Black dots show the measured band powers, with horizontal bars representing the bin widths, and vertical bars representing the uncertainty due to noise, sample variance, and beam uncertainty, the diagonal of the band power covariance matrix.
The red curve is the \wmap-9 \lcdm\ theory, and the red crosses are the expected binned band powers. 
The data are consistent with \lcdm, as described in \secref~\ref{sec:self_calibration}.}
\end{figure}

A single estimate of the power spectra $\hat{C}_b^{XY}$ from the three patches is created using the band powers and their covariance matrices, as will be described in \secref~\ref{sec:analysis}.  
The power spectra are gain-calibrated by fitting the patch-combined \cltt\ to the \wmap-9 \lcdm\ spectrum. 
The patch-combined \cltt, \clee, \clte, and \cltb\ spectra (after the global reference polarization angle calibration using \cleb\ is applied) are plotted in \figref~\ref{fig:spectrumresults}. 
We find that the patch-combined and individual patch spectra are consistent with the \lcdm\ model, where the binned uncertainties on each spectra are from sample variance, noise variance, and beam uncertainty. The patch-combined spectra have a probability-to-exceed (PTE) to the \wmap-9 best-fit \lcdm\ model of 22\%, 54\%, 60\%, 84\%, and 68\% for \cltt, \clee, \cleb, \clte, and \cltb, respectively.

As described in \secref~\ref{sec:polmodel}, \taua\ is used to calibrate the relative pixel polarization angles. 
We use simulations of instrumental systematic effects in \secref~\ref{sec:sys-relpol} to show that our uncertainty in relative pixel polarization angle, and in all other instrumental systematics, does not contribute significantly to the \clbb\ or \cleb\ spectra. 
This allows us to use the \cleb\ spectrum as a more precise calibration of instrument polarization angle to search for the signature of gravitational lensing in \clbb\ \citep{ksy2013}. 
Miscalibration of the instrument polarization angle biases the measured \clbb\ spectrum and produces non-zero \cleb\ and \cltb\ spectra. 
The bias in \clbb\ and non-zero \cleb\ corresponding to an instrument polarization angle error $\Delta \psi \ll \pi$ are given by

\begin{align}
{C_\ell '}^{BB} & \simeq 4 \Delta\psi^2 C^{EE}_\ell \,,\\ 
{C_\ell '}^{EB} & \simeq 2 \Delta\psi C^{EE}_\ell\,.
\label{eq:BBspectra}
\end{align}

A cosmic rotation of polarization would produce a non-zero \cleb\ that is degenerate with an instrument polarization angle miscalibration. 
Either signal can be removed by rotating the instrument polarization angle to minimize the best-fit angle as measured by \cleb\ and \cltb. 
For this analysis, we calibrate the instrument polarization angle using the patch-combined \cleb\ spectrum, which is more sensitive than \cltb\ \citep{ksy2013}. 
We then find consistency between \cltb\ and \cleb, and find that each patch is individually consistent with the single \cleb-defined instrument polarization angle, 
which has a statistical uncertainty of \rotuncert\textdegree. 
Note that this process is expected to minimize the measured \clbb, as any miscalibration of polarization angle or cosmic rotation of polarization increases the power in \clbb \citep{kaufman2014}. 

\begin{figure}[htbp]
 \centering
 \includegraphics[width=3.4in]{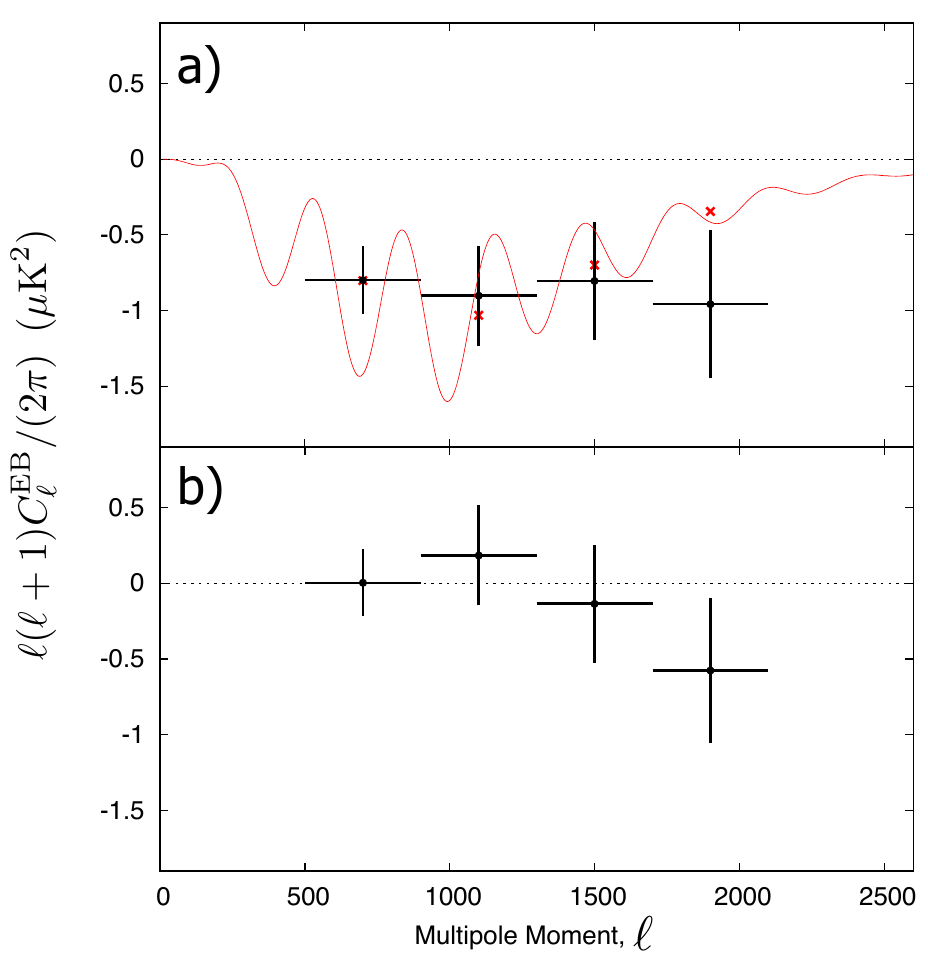}
 \caption{\label{fig:tauaEB}
(a)~The \cleb\ band powers measured using the \taua-derived instrument polarization angle calibration, combining data from all three patches. 
A theoretical $C_{\ell '}^{EB}$ spectrum, expected if the instrument polarization angle calibration is incorrect, is fit to the data (best fit curve shown in red, with binned band powers as red crosses). 
The measured \cleb\ band powers fit the model that the true instrument polarization angle is \rotval\textdegree\ from the \taua-derived instrument polarization angle, with a PTE of 55\%. 
(b)~\cleb\ spectrum, \textit{after} self-calibrating using the \cleb\ data above.
}
\end{figure}

\Figref~\ref{fig:tauaEB} shows the \cleb\ power spectrum measured using the \taua\ calibration of instrument polarization.
This shows that the instrument polarization angle calibrated by \cleb\ is different from the \taua-derived polarization angle by \rotval\textdegree; the statistical uncertainty in the global \cleb-derived instrument polarization angle is \rotuncert\textdegree. 
Given the uncertainty in the IRAM-measured angle of 0.5\textdegree, the \pb\ measurement uncertainty estimated in \secref~\ref{sec:polmodel} of 0.43\textdegree, and the statistical uncertainty of the \cleb-derived angle, there is slight tension between the \taua\ angle measurement and the \cleb\ angle measurement. 
Because of the complicated astrophysics associated with \taua, we believe that \cleb\ is a more accurate measurement of instrument polarization angle to reference to the CMB, with the added benefit that it is more precise. 
We show in \secref~\ref{sec:systematics} that \taua\ is more than sufficient as a relative calibration between pixel angles, because relative uncertainties across the focal plane are mitigated by averaging of many pixels and sky rotation. 
The effect of the \cleb\ statistical uncertainty on \clbb\ is shown in \figref~\ref{fig:total_systematics_BB},
and corresponds to less than $2\%$ contamination of the measured \clbb\ signal.

%%%%%%%%%%%%%%%%%%%%
\section{Power Spectrum Analysis}
\label{sec:analysis}

We estimate angular power spectra using a pseudospectrum technique, based on the MASTER algorithm \citep{Hivon:MASTER}.  Time-ordered data are filtered and binned into flat-sky maps.  Maps are Fourier transformed and the cross spectra between maps form noise bias free power spectrum estimates called pseudospectra.  The pseudospectra are biased by the filtering and partial sky coverage, which we correct for with Monte Carlo simulations and analytical calculations of mode mixing.  Error bars are calculated from the noise in the maps and validated with time-domain noise simulations.

\subsection{Data Selection}
\label{sec:cuts}
We immediately discard all data obtained while the telescope is accelerating, which removes 36\% of our observation time. 
As described in section \ref{sec:obs_yield}, we also exclude some bolometers from ever being used, leaving 746 bolometers which are included in the final set of maps.
Beyond these cuts, we impose additional selection criteria to remove data from consideration where the instrument was not functioning properly.
The selection process can exclude entire days or individual CESs from being processed, remove single pixels from a given CES, or remove individual sub-scans for one pixel, depending on the type of problem detected.

Entire scans are excluded from analysis for low yield, proximity to the sun or moon, high precipitable water vapor, high scan-synchronous signals (see section \ref{sec:ana-filtering}), being in a particular elevation range where the telescope experiences a mechanical resonance, or a malfunctioning telescope encoder.

Pixels are excluded from a single scan for having an outlier gain or differential gain value or if either of these quantities changes too rapidly over the course of the scan. 
Pixels are also removed for high noise or unphysically low noise, as well as having noise higher in one subscan direction than the other.

We also see features in the timestreams -- glitches -- for which we are unsure of the source. 
Subscans that show these glitches are simply removed from the data set. 
To flag subscans as containing glitches, we convolve differenced timestreams with a set of Lorentzian-based kernels and remove subscans which exceed seven times the median absolute deviation of the convolved timestream. 
There are other flagging criteria imposed on subscans where there were problems with the bolometer or telescope pointing data acquisition systems.

The data selection criteria described here, focusing on the constant-velocity portions of the CESs, remove 40\% of the data by weight.  

\subsection{Time Ordered Data to Maps} 
\label{sec:mapmaking}

Following \cite{chiang2010}, we construct maps $m$ from detector time streams $d$, diagonal detector variance estimates $N$, a time-domain filter $F$, and pointing matrix $A$, using

\begin{equation} (\mathbf{A}^T \mathbf{N}^{-1} \mathbf{A}) \mathbf{m} = \mathbf{A}^T \mathbf{N}^{-1} \mathbf{F} \mathbf{d}\,. \label{mapmakingeqn} \end{equation}
This is a noise weighted, biased estimate of the sky signal. The procedure is described in more detail below.

\subsubsection{Low-level data processing}

Bolometer TOD are low-pass filtered with an anti-aliasing filter matched to the digitization sample rate of 190.7\,Hz.
The TOD-based data selection described in \secref~\ref{sec:cuts} is then applied.
The telescope azimuth and elevation encoders are queried at 95.4\,Hz, synchronous but offset with every other sample of the bolometer timestream. 
The telescope pointing is reconstructed using the procedure described in \secref~\ref{sec:pointing} and linearly interpolated to the bolometer sample times.
Because this analysis focuses on $\ell<2100$, corresponding to time domain frequencies less than 4.2\,Hz, we downsample the bolometer TOD and reconstructed pointing by a factor of six to 31.8\,Hz.
The bolometer optical response times, which are between 1--3\,ms, are small enough that deconvolving these transfer functions is not necessary \citep{Arnold_SPIE2012}.

Individual bolometer timestreams are calibrated as described in \secref~\ref{sec:gain}, then pixel-pair bolometers are summed and differenced to derive temperature and polarization timestreams from each pixel.  The polarization timestreams contain a linear combination of Stokes $Q$ and $U$ signal depending on the polarization orientation of the detector projected onto the sky. 

\subsubsection{Filtering}
\label{sec:ana-filtering}

Timestreams are noise weighted with a filter $F$ which projects out three types of low signal-to-noise-ratio modes:  High frequencies, low order polynomials per subscan, and scan synchronous signals.

Because map making, pixelized at a Nyquist frequency of $\ell=5400$, is a decimation operation from the timestreams, pixelized at a Nyquist frequency of $\ell=7950$, high frequencies can alias into our science band.  
We apply a convolutional low-pass filter to the timestreams to eliminate aliasing.  
The frequency profile of the filter is

\begin{equation}
F_{\rm lpf}(f) = e^{-\log(\sqrt 2) \left( \frac{f}{f_{\rm lpf}}\right)^6}\, ,
\end{equation}
where the 3\,dB frequency of the filter is $f_{\rm lpf} = 6.3$\,Hz, corresponding to an angular multipole of $\ell=3150$.

To remove excess low-frequency noise, for each detector we subtract a polynomial per CES subscan.
Linear polynomials are used for difference (polarization) timestreams, and cubic polynomials for sum (temperature) timestreams.
Point sources (described further in \secrefs~\ref{sec:pdbeams} and \ref{sec:fg}) are masked with a 10\arcmin\ diameter mask during fitting to keep the point source power localized.  
The exception is Mars, which traverses the RA12 patch multiple times during the season.  Because of Mars's brightness, TOD are masked within a 30\arcmin\ diameter disk centered on it during the polynomial filtering, and this masked data is also excluded from the maps.

Our scan strategy is designed to concentrate scan-synchronous signals, such as a far sidelobe scanning the ground, into a small number of modes which can be easily filtered.
During a CES, scan-synchronous signals will repeat in azimuth for the duration of the scan.
These signals are projected out by averaging the timestreams in $0.08^\circ$ azimuth bins for each bolometer to build a scan-synchronous signal template; the template is then subtracted from the timestreams.
Because the CMB patch moves by about $3^\circ$ in this time, the effect of this filter on sky signal has only a small impact on the multipole range reported here, $500 < \ell < 2100$, as shown in \figref~\ref{fig:TransferFunction}.

\subsubsection{Pointing matrix \& noise weights}

The pointing matrix and noise weights used in Equation~\ref{mapmakingeqn} are split into components acting on the sum and difference timestreams of the two orthogonally polarized detectors in one focal plane pixel.
The pointing matrix for the sum component maps only Stokes intensity $I$ from the sky into the detector timestream $d_t^{\rm sum}$, 

\begin{equation} d^{\rm sum}_t = \mathbf{ A }_{tp} I_p + n^{\rm sum}_t. \end{equation}

\noindent The pointing matrix for the difference timestream is assumed to have no temperature response and only Stokes $Q$ and $U$ response,

\begin{equation} d^{\rm diff}_t = \mathbf{ A }_{tp} \cos(2 \psi) Q_p + \mathbf{ A }_{tp} \sin(2 \psi) U_p + n^{\rm diff}_t. \end{equation}
The assumption of no temperature response is described in detail in Section \ref{sec:systematics}.
Detector pointing is projected onto a flat map using the cylindrical equal area projection centered at the nominal patch center \citep{USGSProjections}. 
The map pixels have a width of 2\arcmin.

\label{sec:noise_weights}

The noise weights $\mathbf{N}^{-1}$ used for co-adding the pixel maps are estimated under an idealized model that the noise is white.  The time domain power spectral density are averaged from 1--3\,Hz to estimate the detector noise variance.  This frequency band corresponds approximately to $\ell$ of 500--1500.

\subsubsection{Daily maps}
\label{sec:daily_maps}

We co-add the single-CES, all-detector maps into a set of single-day $I$, $Q$, and $U$ maps.  
The patches RA4.5, RA12, and RA23 have 148, 139, and 189 daily maps respectively.
The map making procedure produces inverse noise covariance estimates for each pixel describing the covariance of Stokes $I$, $Q$, and $U$.
Polarization maps are apodized using the minimum eigenvalue of the $2\times 2$ $Q$ and $U$ block of the pixel noise inverse covariance matrix.  
This eigenvalue is an upper bound for the variance in either $Q$ or $U$.
Temperature maps are apodized with an inverse noise variance apodization at the edge of the map, and with a flat function where the CMB fluctuations are measured at a large signal to noise ratio.

Pixels with an apodization window value below 1\% of its peak are set to zero, as are pixels within 3\arcmin\ of point sources.
To reduce $E/B$ leakage, the apodization edges are modified using the $C^2$ taper described in \citet{xpure}, which is applied inward with a 30\arcmin\ width.

\begin{figure*}[htpb]
 \centering
 \includegraphics[width=6.5in]{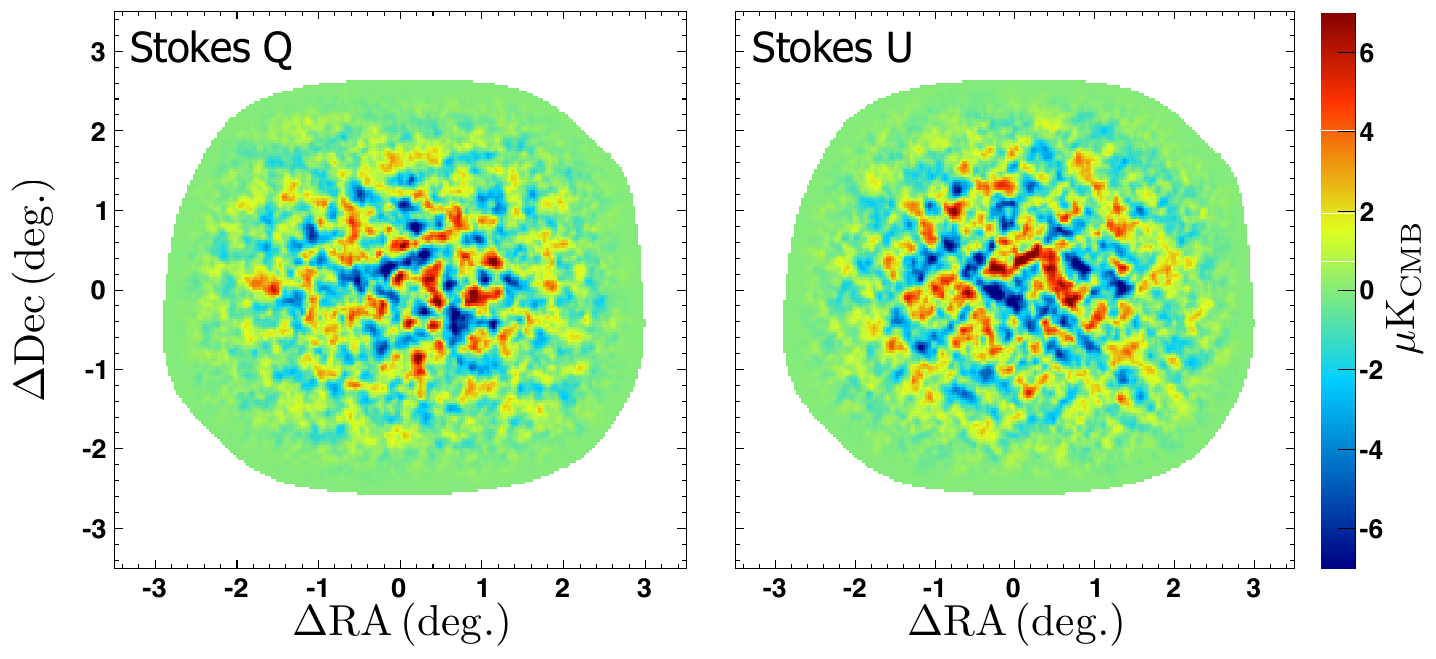}
 \caption{\pb\ CMB polarization maps of RA23 in equatorial coordinates.
 The left~(right) panel shows Stokes $Q$~($U$), where the polarization angle is defined with respect to the North Celestial Pole. 
These filtered maps are smoothed to 3.5\arcmin\ FWHM. 
The clearly visible coherent vertical and horizontal patterns in the $Q$ map and diagonal patterns in the $U$ map are the expected signature of an $E$-mode signal.}
 \label{fig:QUmaps}
\end{figure*}

$Q$ and $U$ maps are transformed to $E$ and $B$ maps. 
The $B$ maps are estimated from the pure $B$-mode transform \citep{Smith2006}.  
Before Fourier transforming, the maps are padded to a width of 384 pixels or $12.8^\circ$ to eliminate spurious periodic effects.
Co-added maps from the entire season are not used in the analysis; however we show them here in \figref~\ref{fig:QUmaps} for $Q$ and $U$ on RA23, which results in a polarization white noise level of 6\,\uK-arcmin~(8\,\uK-arcmin with the beam and filter transfer function divided out.
See \secref~\ref{sec:beammodel} and \secref~\ref{sec:powerspectrumestimation}).

\subsection{Power spectrum estimation}
\label{sec:powerspectrumestimation}

%Put the below at beginning of analysis section

%Reduction of Polarbear data is accomplished through the MASTER framework \cite{Hivon:MASTER}:  Filtered map making followed by pseudo power spectrum estimation, which is de-biased using monte carlo simulations.

%\subsubsection{Power spectra from daily maps}

Six pseudo-power spectra $\tilde{C}_\ell$ ($TT$, $EE$, $BB$, $TE$, $TB$, $EB$) are formed by taking cross spectra of the apodized and Fourier transformed single-day maps $\tilde m^X_{i k}$ for $X \in T$, $E$, or $B$, and day $i$.  
This estimator is free of noise bias \citep{hinshaw2003}.  
The two-dimensional cross-spectra are binned by Fourier mode $k$ in rings of width $\Delta k=40$ to form one-dimensional spectra,

\begin{equation} \tilde C^{XY}_\ell = \frac{1}{\sum_{i,j \neq i,k \in {\rm bin}_\ell} w_i^X w_j^Y} \sum_{i,j \neq i, k \in {\rm bin}_\ell} w_i^X \tilde{\mathbf{m}}_{ik}^X w_j^Y \tilde{\mathbf{m}}_{jk}^{Y*} .
\end{equation}
The weights for the maps in the cross-spectrum procedure, $w_i^X$, are the sum of the pixel inverse noise covariance estimate over all the map pixels, either the $TT$ element for temperature or the minimum eigenvalue of the $Q$ and $U$ block for polarization.

The map making and pseudo-power spectrum procedure are modeled as a linear function of the true sky power spectra $C_\ell$:

\begin{eqnarray} 
\tilde C_\ell = \sum_{\ell'} \mathbf{K}_{\ell\ell'} C_{\ell'}, \label{MASTEREQN} \\
\mathbf{K}_{\ell\ell'} = \mathbf{M}_{\ell\ell'} F_{\ell'} B_{\ell'}^2.
\end{eqnarray}
\noindent \Mll\ describes the mode mixing effects of non uniform sky coverage, and is calculated analytically.
$F_{\ell'}$ models the transfer function of the time-domain filters and map pixelization, and is calculated through \mc\ simulations.
$B_{\ell'}$ describes the smoothing due to the spatial response of the detector.

\subsubsection{Mode-mixing and filter transfer functions}
\label{sec:mll_and_fl}

\Mll\ is computed analytically, by co-adding the temperature and polarization apodization windows from the daily maps for the entire season.
The resulting window map is used to calculate $\mathbf{M}_{\ell\ell'}$ \citep{louis2013}. 

We estimate the transfer function $F_\ell$ of the time domain filters from a suite of \mc\ simulations.
The input to the \mc\ simulations is a set of 1\arcmin-resolution Gaussian realizations of a $10^\circ \times 10^\circ$ patch of the CMB from the best fit \wmap-9 power spectra, $C_{\ell}$ \citep{2013ApJS..208...20B}.
We use the pointing data from observations to produce TOD from the simulated maps, and apply the pseudo-power spectrum estimation procedure. 
We then estimate the filter transfer function from

\begin{equation}
F_\ell^n = F_\ell^{n-1} + \frac{\tilde C_\ell - \sum_{\ell'} \mathbf{M}_{\ell\ell'} F_{\ell'}^{n-1} C_{\ell'} B_{\ell'}^2}{C_\ell B_\ell^2}, \label{filtersolve}
\end{equation}

%\begin{equation}
%F_\ell = 1 + \frac{\tilde C_\ell - \sum_{\ell'} \mathbf{M}_{\ell\ell'} C_{\ell'} %B_{\ell'}^2}{C_\ell B_\ell^2}. \label{filtersolve}
%\end{equation}
\noindent with $F_\ell^0 = 1$, and convergence achieved within the 10 iterations used to calculate $F_\ell = F_\ell^{10}$.

\begin{figure}[htpb]
 \centering
 \includegraphics[width=3.2in]{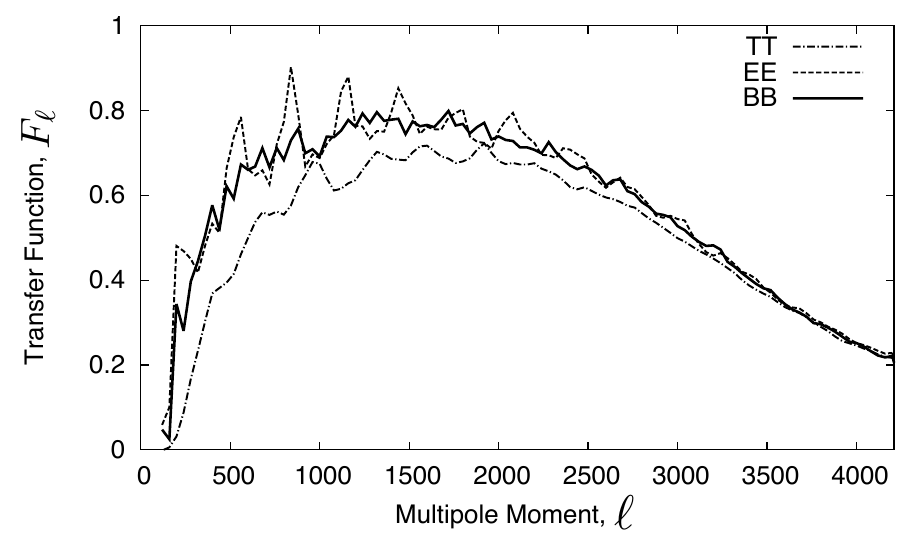}
 \caption{\label{fig:TransferFunction}
Filter transfer functions $F_\ell$ for $TT$, $EE$, and $BB$ power spectrum estimators, calculated from \mc\ simulation. 
The decrease at low $\ell$ is due to the first and third order polynomial filtering of subscans in polarization and temperature respectively. 
The decrease at high $\ell$ is due to the 6.3\,Hz low-pass filtering and the pixelization of the maps \citep{Wu_2001}. 
The structure in the $EE$ transfer function reflects some residual dependence on cosmology in the transfer function calculation. 
It has negligible impact on the \clbb\ result reported here.
}
\end{figure}

To distinguish between leakage and transfer function effects, the filter transfer functions for $E$ and $B$ are computed from separate $TT+EE$ and $TT+BB$ simulations.
The $TE$, $TB$, and $EB$ spectra filter transfer functions are estimated as the geometric mean of the respective auto spectra.
$TT$, $EE$, and $BB$ transfer functions are shown in \figref~\ref{fig:TransferFunction}.

Polynomial filtering and scan-synchronous signal subtraction create leakage from \clee\ to \clbb.
The leakage transfer function is estimated from the \mc\ simulations and leakage is subtracted in power spectrum estimation.
Equation~\ref{filtersolve}, $TT+EE$ simulations, with the $EE$ theory for $C_l$ and $BB$ pseudospectra and mode mixing matrix are used to estimate $F^{E \to B}_\ell$.
Before subtraction, the leakage is largest in the lowest bin centered at $\ell=700$ where it is 9\% of the \clbb band power.
The power is subtracted in pseudospectrum space with an amplitude of
\begin{equation}
 \tilde C^{E \to B}_\ell = \frac{F^{E \to B}_{\ell}}{F^{E \to E}_{\ell}} \tilde C^{E}_\ell.
\end{equation}
The uncertainty associated with this subtraction is calculated via \mc\ simulations that include $TT$, $EE$, and $BB$ power. The residual bias and its uncertainty, including sample variance, is calculated as $\ell(\ell+1)C_\ell^{BB}/(2\pi) = (6.1\pm 39.9)\times 10^{-4}$\,\uKsq\ at $\ell=700$.
This uncertainty is included in the presented limits on \ABB.

\subsubsection{Band power window functions}

$F_\ell$ and $\mathbf{M}_{\ell\ell'}$ are calculated with a resolution of $\Delta \ell=40$.
We solve for the unbiased estimates of the sky spectra on four coarser bins of width $\Delta \ell=400$ within $500 < \ell < 2100$ using binning and interpolation operators $\mathbf{P}_{b\ell}$ and $\mathbf{Q}_{\ell b}$.
The unbiased, binned estimator for the $\Delta \ell=400$ binned power spectra is

\begin{eqnarray} 
\hat C_b = \sum_{b'\ell} \mathbf{K}^{-1}_{bb'} \mathbf{P}_{b'\ell} \tilde C_{\ell}, \label{MASTERSOLN} \\
\mathbf{K}_{bb'} = \sum_{\ell\ell'} \mathbf{P}_{b\ell} \mathbf{M}_{\ell\ell'} F_{\ell'} B_{\ell'}^2 \mathbf{Q}_{\ell'b'}.
\end{eqnarray}

\begin{figure}[htpb]
 \centering
 \includegraphics[width=3.2in]{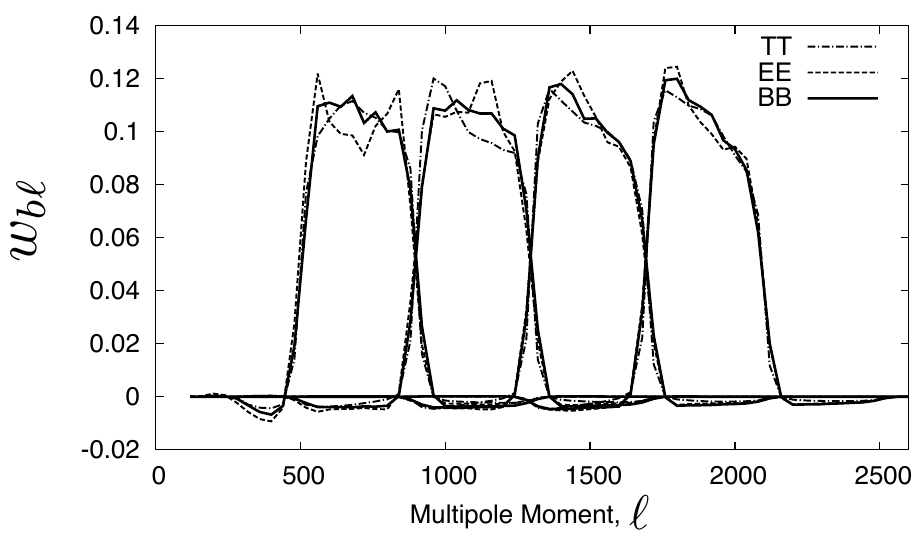}
 \caption{Band power window functions $w_{b\ell}$ describing the transfer function for $TT$, $EE$ and $BB$ power
 from a $\ell(\ell+1)C_\ell / 2\pi$ power spectrum to binned band powers,
 as described in equation~\ref{eqn:bpwf}.}
 \label{fig:BPWF} 
\end{figure}

The functional dependence of the binned band powers $\hat C_b$ on the true high-resolution power spectra is given by the band power window functions $\mathbf{w}_{bl}$, where

\begin{eqnarray}
\hat C_b = \sum_{l} \mathbf{w}_{b\ell} C_\ell , \\
\mathbf{w}_{bl} = \sum_{b'\ell'} \mathbf{K}^{-1}_{bb'} \mathbf{P}_{b'\ell'} \mathbf{K}_{\ell'\ell} \label{eqn:bpwf}.
\end{eqnarray}
The resulting band power window functions for temperature and polarization are shown in \figref~\ref{fig:BPWF}. 

\subsection{Power Spectrum Uncertainty Estimation}
\label{sec:power_spec_error_bars}

Uncertainty in the binned power spectrum is analytically estimated.
For $XX \in TT, EE$, and $BB$, the uncertainty estimate is
\begin{equation}
\Delta \hat C_b^{XX} = \sqrt{\frac{2}{\nu_b^{XX}}} (C_b^{XX} + \hat N_b^{XX}).
\end{equation}
For $XY \in TE, TB$, and $EB$, the uncertainty estimate is
\begin{equation}
\Delta \hat C_b^{XY} = \sqrt{\frac{(C_b^{XY})^2 + (C_b^{XX} + \hat N_b^{XX})(C_b^{YY} + \hat N_b^{YY} )}{\sqrt{\nu_b^{XX} \nu_b^{YY}}}}. \label{eqn:crosserrbars}
\end{equation}
The sample variance term $C_b$ is the binned \wmap-9 \lcdm\ spectra. 
The binned noise spectrum $\hat N_b$ is estimated from the auto spectrum $\hat A_b$ of the fully coadded map, corrected for the cross spectrum estimate of the signal,

\begin{equation} \hat N^{XX}_b = \hat A^{XX}_b - \hat C^{XX}_b ,\end{equation}

\begin{align}
\hat A^{XX}_b  = \sum_{b'\ell} & [ \mathbf{K}^{-1}_{bb'} \mathbf{P}_{b'\ell} \frac{1}{\sum_{i,k \in {\rm bin}_\ell} w_i^X w_i^X} \nonumber \\
  & \times \sum_{i,k \in {\rm bin}_\ell} w_i^X \tilde{\mathbf{m}}_{ik}^X w_i^X \tilde{\mathbf{m}}_{ik}^{X*} ].
\end{align}

The $\nu_b$ term is the effective number of degrees of freedom in each bin $b$.
The second and fourth moments $w_2$ and $w_4$ of the apodization window averaged over the daily maps for the entire season is used to calculate $\nu_b$, 

\begin{equation}\nu_b = (2\ell_b + 1) \Delta \ell f_{\rm sky} \frac{w_2^2}{w_4},
\end{equation}

\noindent where the term $ f_{\rm sky}{w_2^2}/{w_4}$ is the effective sky area for each patch.

This binned uncertainty estimation is validated using full-season Monte Carlo simulations including signal and noise. 
We chose two noise models on which to test this estimator, a white noise model and a correlated noise model. 
In each case, the spread in the \cleb\ and \clbb\ power spectra obtained from the \mc\ simulations is consistent with the mean result of the analytic binned uncertainty estimator to 10\%. 

The simulations use the same pointing and detector weighting as the real data, and include signal from a beam-convolved realization of a \wmap-9 \lcdm\ power spectrum that includes the effect of gravitational lensing.
The white noise model adds random spectrally flat noise to the timestream of each detector variance equivalent to that measured from the detectors as described in \secref~\ref{sec:noise_weights}.

To test our binned uncertainty estimator using correlated noise, we consider every TOD $d_i^{\ (t,\; S)}$ from detector $i$, over a group of several consecutive subscans, indexed by $S$. 
Each  timestream is then apodized with a Hanning window and Fourier transformed, resulting in vectors $d_{i}^{\ (f,\; S)}$. 
Similarly to what was proposed by \cite{chiang2010}, this vector is binned in frequency space, and the full binned covariance matrix estimator is defined as
\begin{eqnarray}
	\centering
		\mathbf{C}_{ij}^{\ b} \equiv \left\langle \left\langle  d_{i}^{\ (f,\; S)}  d_{j}^{\ (f,\; S)^*} \right\rangle_{f\,\in\, b} \right\rangle_{S}.
	\label{eq:Cijb_def}
\end{eqnarray}
Assuming the noise is uncorrelated between CESs, a unique $\mathbf{C}_{ij}^{\ b}$ is estimated for each of them. $\mathbf{C}_{ij}^{\ b}$ is a complex object of size $\left( n_{det} \times n_{det} \times n_{bins} \right)$ which satisfies
\begin{eqnarray}
	\centering
		\forall\ b,\ \mathbf{C}_{ij}^{\ b} = \left( \mathbf{C}_{ji}^{\ b}\right)^* .
\end{eqnarray}
We use groups of 25 consecutive subscans for $S$, and 12 logarithmically-spaced bins between 0.001 and 15\,Hz for $b$.
Simulated noise TOD for each detector and each CES are generated as a random realization of the binned covariance matrix $\mathbf{C}_{ij}^{\ b}$, and the angular power spectra are calculated from the resulting simulated maps. 

For both noise models, the uncertainty estimator correctly recovers the spread in the power spectra estimated from random realizations of the sky and the noise to within 10\%.  In polarization, this indicates that the difference timestream noise is very white.  In temperature, the sum timestream noise is not white, but the signal-to-noise is extremely high and the binned uncertainty is dominated by sample variance in the entire multipole range reported here.

%%%%%%%%%%%%%%%%%%%%
\section{Foregrounds}
\label{sec:fg}

Four foreground sources of polarized emission are of potential concern at the frequency and angular scale of the reported measurement. 
On large scales, polarized galactic dust and synchrotron emission are most important, while on smaller scales, polarized radio and dusty galaxies are most important. 
The expected contamination by galactic foregrounds has been mitigated by the careful selection of the three \Pb\ fields, which were chosen primarily to reduce the contamination due to the polarized galactic dust emission.
The foregrounds are summarized in \tabref~\ref{tab:foregrounds}. 
The following paragraphs describe how this table was derived. 

\begin{table}
\begin{center}
\caption{\label{tab:foregrounds}Sources of foreground power and their predicted power in \dlbb. All foreground power is small compared to our statistical uncertainties, and we do not subtract it. The total bias on the final line is the linear sum of the individual foreground powers.}
\begin{tabular}{ p{1.1in} | p{0.4in} | p{0.4in} | p{0.4in} | p{0.4in}}
\hline
\hline
Foreground & \multicolumn{4}{l}{ Predicted power in \dlbb }  \\
 & \multicolumn{4}{l}{($10^{-4}$ \uKsq)} \\
\hline
 & 500-  900 & 900-1300 & 1300-1700 & 1700-2100 \\ 
\hline
Galactic dust & 40 & 28 & 22 & 9.7 \\
Galactic synchrotron & 2.3 & 1.8 & 1.6 & 1.4 \\
Radio galaxies & 6.3 & 15 & 28 & 46 \\
Dusty galaxies & 2.8 & 4.5 & 6.5 & 8.7 \\
\hline
Total bias & 51 & 49 & 58 & 66 \\
\hline
\end{tabular}

\end{center}
\end{table}

The expected level of dust contamination is computed from the publicly available Planck Sky Model (PSM) software described by \citet{Delabrouille_2013}. 
For this purpose, we only consider thermal dust emission; other diffuse dust components, such as anomalous microwave emission, are expected to be subdominant at the \pb\ frequency of observation \citep{Planck2013XII}. 
We conservatively multiply the PSM templates by a factor of two (increasing the polarization power spectrum by a factor of four) given the comparatively larger polarized fraction recently reported in other parts of the sky at 353\,GHz \citep{Planck2014XIX}. 
We compute the angular power spectrum of the dust emission in each patch using the actual \pb\ masks and pixel weights. 
The final dust band powers are a noise-weighted average of the dust power spectrum for each patch.

\citet{quiet_SecondSeason} measured the polarized Galactic synchrotron power at $\ell=50$ and 95\,GHz to be less than $0.005$\,\uKsq{} at $2\sigma$ in the deepest two \Pb{} fields. 
This power is already quite low and galactic synchrotron power decreases at smaller angular scales and higher frequencies. 
We use a spectral dependence of $\nu^{-2.7}$ to scale to the \pb\ observing frequency \citep{Dunkley2009}. 
We use the angular scale dependence $C_\ell \propto \ell^{-2.5}$ \citep{laporta08} to apply the \quiet\ measurements to the band powers reported here.

The small \pb\ beam makes it possible to detect and remove point sources -- in their deepest regions, the \pb\ maps have a $5\sigma$ source detection threshold of 25\,mJy, a factor of about seven below that of \planck\ at the same frequency \citep{Planck2013XXVIII}. 
We mask out sources above 25\,mJy, which is the $5\sigma$ detection threshold over the deepest 60\% of the area used in this analysis. 
These sources are masked when calculating the polynomial timestream filters (see \secref~\ref{sec:ana-filtering}), then they are masked in daily maps before power spectrum estimation (\secref~\ref{sec:daily_maps}). 
All of the sources we detect correspond to sources detected by either ATCA \citep{at20g} or Planck \citep{Planck2013XXVIII}. 
The unmasked point sources below the 25\,mJy detection threshold contribute a residual $TT$ power of 6.2\,\uKsq\ at $\ell=3000$.
This estimate was derived by calculating the power in simulated point source maps based on number counts of galaxies from \citet{dezotti05} obtained through personal communication with the authors of that paper.
The polarization fraction has been measured to be 0.01--0.05 at 20\,GHz \citep{sadler06}, we use the upper bound of 0.05 to estimate the polarized radio galaxy power. 

Although we expect slightly more $TT$ power from dusty galaxies than radio galaxies,  their polarization fraction should be lower. 
We conservatively use the 99\% confidence upper limit of 1.54\%  polarization reported by \citet{seiffert07} as the RMS polarization fraction, and use the measured temperature anisotropy power from dusty galaxies \citep{reichardt12}. 
Using the \spt-derived dust power is conservative since the \pb\ band centers are slightly lower than SPT's 153.8\,GHz band center.
Under these assumptions, the polarized power from dusty galaxies is

\begin{align}
\ell(\ell+1)C_\ell^{BB,dg} = 0.5 \cdot & 0.0154^2 \cdot  [ 7.54 \left ( \ell/3000 \right )^2 \notag \\
 &  + 6.25 \left ( \ell/3000 \right )^{0.8}  ] (\mu \mathrm{K}^2).
\end{align}

%%%%%%%%%%%%%%%%%%%%
\section{Systematic instrumental effects}
\label{sec:systematics}

This section describes our evaluation of spurious instrumental effects on \clbb. %, and shows that they are small compared to the measured $B$-mode angular power spectrum. 
\Tabref~\ref{tab:sys-bias} summarizes the systematic uncertainties we considered the most important possible sources of bias in the measurement of \cleb\ and \clbb, and \tabref~\ref{tab:sys-calib} outlines those affecting the calibration of the \cleb\ and \clbb\ signal. 
The absolute calibration uncertainty is calculated from the fit to \wmap-9 \cltt, the uncertainty in the polarization efficiency described in \secref~\ref{sec:polmodel}, and an analysis of the sensitivity of $F_\ell^{BB}$ to different \clbb\ cosmologies as input to the \mc\ simulations from which $F_\ell$ is calculated. 
The fractional calibration uncertainties from each of these studies are then added in quadrature to compute a total symmetric calibration uncertainty of 6.7\%.

We apply three frameworks for the investigation of instrumental bias of the \cleb\ and \clbb\  power spectrum. Signal-only simulations used to determine the effect of  instrument model uncertainties on the power spectrum are described in \secref~\ref{sec:sys-sim}; special analyses of the data focused on illuminating possible effects of instrumental contamination are described in \secref~\ref{sec:sys-special}; null tests to show that the data set is internally consistent are described in \secref~\ref{sec:sys-null}. 
\Secref~\ref{sec:ubpipe} describes an alternate pipeline we have been developing that was used as a cross-check of these results.
\Secref~\ref{sec:blind} describes the blind analysis strategy we employed to mitigate any effects of observer bias.
The result of these analyses is that none of the instrumental effects taken into account produce significant contamination of the \Pb\ \clbb\ measurement. 
The calculated upper bound on the sum of all considered systematic contamination in \clbb\ is shown in \figref~\ref{fig:total_systematics_BB}. 
To evaluate the effect of this systematic uncertainty on the measurement, these binned upper bound values are conservatively added linearly together with the binned upper bounds on foreground contamination given in \tabref~\ref{tab:foregrounds}.
Those values are then subtracted from the measured \clbb, %and \ABB\ is fit to these modified band powers. 
and the significance with which we reject the null hypothesis is calculated using these reduced band powers, combined with their statistical uncertainties.
The \ABB\ fit to these reduced band powers sets the lower bound of the reported asymmetric systematic uncertainty on the measured \ABB.

\begin{table*}
\begin{center}
\caption{\label{tab:sys-bias}Estimates of the maximum contribution to \clbb due to instrumental uncertainties that could bias (additively) the $B$-mode signal as estimated using the simulations described in \secref~\ref{sec:sys-sim}. 
The linear sum of these effects in each band power is taken as an upper limit on the possible instrumental bias on the measurement.}
\begin{tabular}{ p{2.25in}  p{2.75in}  p{1.1in} }
\hline
\hline
	\begin{center}Source of uncertainty\end{center} & \begin{center}Measurement technique\end{center} & \begin{center}Maximum spurious \dlbb\ [$10^{-4}$\,\uKsq]\end{center} \\
	\hline
	Boresight pointing  & Comparison of pointing models & 5.5 \\ 
	Differential pointing  &  Planet beam maps  & 7.1 \\ 
	Instrument \& relative polarization angle & \cleb\ statistical uncertainty and \taua & 12 \\
	Pixel-pair relative gain: HWP-independent & Comparison with \taua, differential-gain map-making & 2.2 \\
  	Pixel-pair relative gain: HWP-dependent & Comparison with elevation nods & 9.4 \\
	Pixel-pair relative gain: drift & Comparison of compensation versus no compensation & 0.41 \\
	Differential beam ellipticity & Planet beam maps & 3.3 \\
	Differential beam size &  Planet beam maps & 8.3 \\
	Electrical crosstalk & Simulation of measured level & 1.7 \\
\hline
	Total possible bias & bin central $\ell$: 700, 1100, 1500, 1900 \ & 40, 41, 39, 29 \\
\hline
%	\tablecomments{Benchmarks and estimated values used for this analysis for the different systematics effects.}
\end{tabular}

\end{center}
\end{table*}

\begin{table*}
\begin{center}
\caption{\label{tab:sys-calib}Summary of instrumental and analysis uncertainties that affect the amplitude of the measured $B$-mode signal in a multiplicative way. 
%The table provides how they are measured and their effect on recovering \clbb. 
These are added in quadrature to determine the total calibration uncertainty of the measurement.}
\begin{tabular}{ p{2.25in}  p{2.75in}  p{1.1in} }
                  \hline
\hline
	 \begin{center}Effect\end{center} & \begin{center}Measurement technique\end{center} & \begin{center}Uncertainty in \dlbb \end{center}  \\
	\hline
	 Statistical variance and beam co-variance & \wmap-9 comparison, compact sources & 4.1\% \\ 
	 Polarization efficiency  & HWP model \& \taua  & 3.6\% \\
	 Transfer function & \mc\ varying \clbb\ cosmology & 3.9\% \\
\hline
\end{tabular}
\end{center}
\end{table*}

\subsection{Simulations of instrumental effects}
\label{sec:sys-sim}

All of the instrumental effects in \tabref~\ref{tab:sys-bias} were analyzed using signal-only simulations to highlight the effect of specific instrumental uncertainties. 
The instrumental effects that did not involve beam asymmetries were analyzed using the high-resolution simulation pipeline described below, while the uncertainties due to differential beam properties relied on simulations focusing on CMB-gradient maps, described in \secref~\ref{sec:sys-beams}.
This pipeline is fairly general with respect to the types of systematic errors it can simulate, and is particularly well-suited to effects that involve small deflections in map-space. 

$12^\circ \times 12^\circ$ maps with $3$\arcsec resolution pixels are created from realizations of the theoretical unlensed \lcdm\ spectra multiplied by the symmetric \Pb\ $B_\ell^2$.
These realizations contain $TT$, $TE$ and $EE$ temperature and polarization power, but no $BB$ power.
The maps are scanned with the actual \Pb\ pointing and the instrumental effects in question are injected directly into the timestreams on the fly. $I$, $Q$ and $U$ maps are then reconstructed at the standard \Pb\ map resolution.  
No filtering is included in this process.
The power spectra of these maps are then estimated using the \textsc{x$^2$pure} method \citep{xpure, Grain2012, ferte2013},
which implements the pure-pseudospectrum technique \citep{Smith2006, SmithZaldarriaga2007} to minimize the effects of the $E$-to-$B$ leakage due to the cut-sky effects.
As a result of this framework, any non-zero \cleb\ or \clbb\ power is spurious, and a measurement of the instrumental systematic effect. 
This pipeline, which co-adds daily observations and then auto-correlates to measure power spectra, is slightly more sensitive to some systematic errors than the primary pipeline, which cross-correlates data from different days.
Of course, instrumental effects could also distort an existing \clbb\ spectrum, and these effects could be understood using the pipeline described above by including $B$-mode power in the simulated maps. 
We expect these effects to be small, given the already faint $B$-mode signal. 
In the future, these effects will have to be understood to precisely characterize \clbb. 
Given the statistical uncertainties reported here, we chose not to investigate effects distorting \clbb\ for this study. 

\begin{figure}[htpb]
\begin{center}
\includegraphics[width=3.5in]{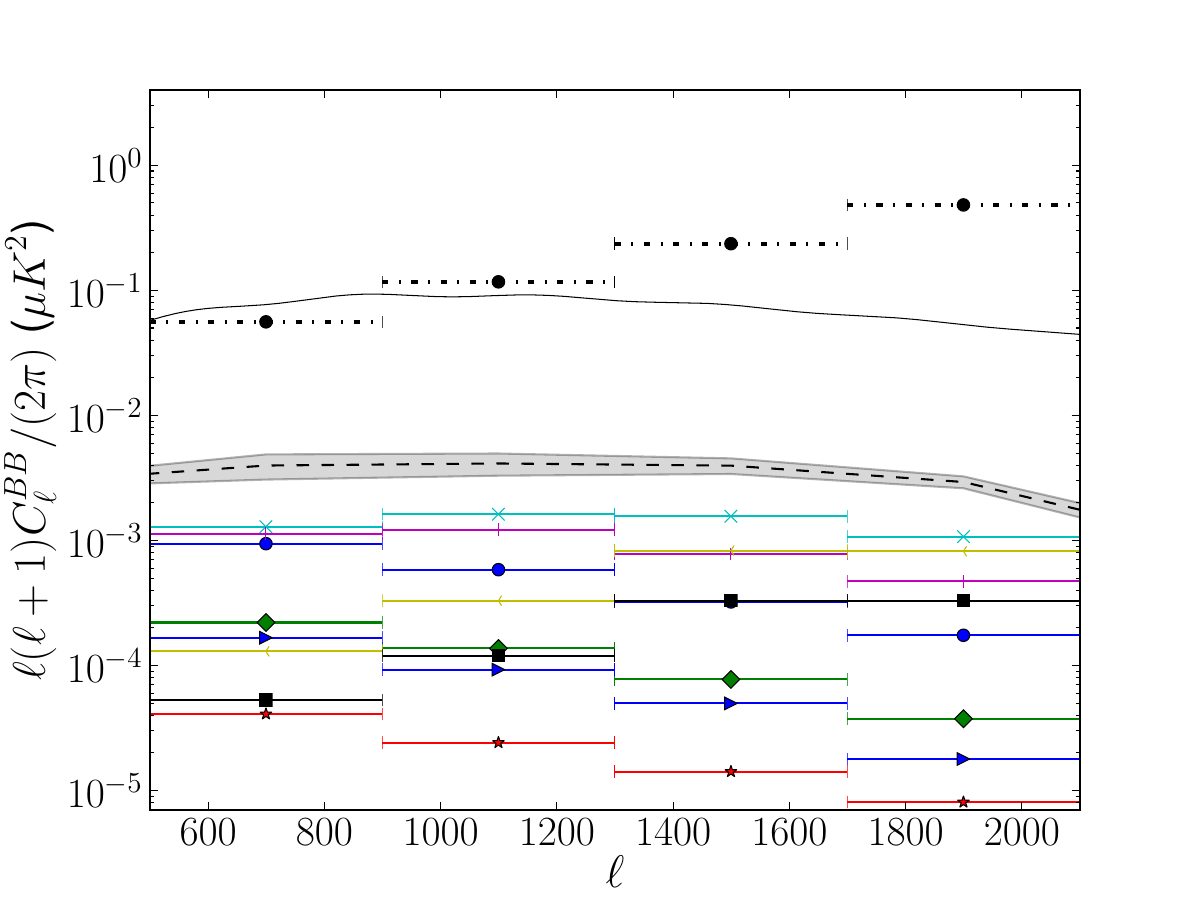}
\end{center}
\caption{\label{fig:total_systematics_BB}
Estimated levels or upper bounds on instrumental systematic uncertainties in the \clbb\ power spectra, as described in \secref~\ref{sec:systematics}. 
Both the individual sources of uncertainty (solid color) and the cumulative bias coming from their combination (black dashed) are displayed after the combination of all CMB patches.
The grey-shaded region show the $1\sigma$ bounds on the cumulative bias limit, after the self-calibration procedure described in \secref~\ref{sec:self_calibration}. 
This is found through \mc\ simulations of our observations with the systematics included.
The effects included in this analysis were the boresight and differential pointing uncertainty (light blue cross mark), the residual uncertainty in instrument polarization angle after self-calibration (purple plus mark), the differential beamsize and ellipticity (yellow arrow and black square mark respectively), the electrical crosstalk (blue arrow mark), the drift of the gains between two consecutive thermal source calibrator measurements (red star mark), and the HWP-independent and HWP-dependent terms of the relative gain model (green diamond and blue circle mark respectively). 
Also shown are the theoretical unlensed \lcdm\ \clbb\ (solid black line) and the binned statistical uncertainties reported in \tabref~\ref{tab:bandpowers} (black bullets with horizontal bars)
}
\label{fig:BB-systematics}
\end{figure}

\begin{figure}[htpb]
\begin{center}
\includegraphics[width=3.5in]{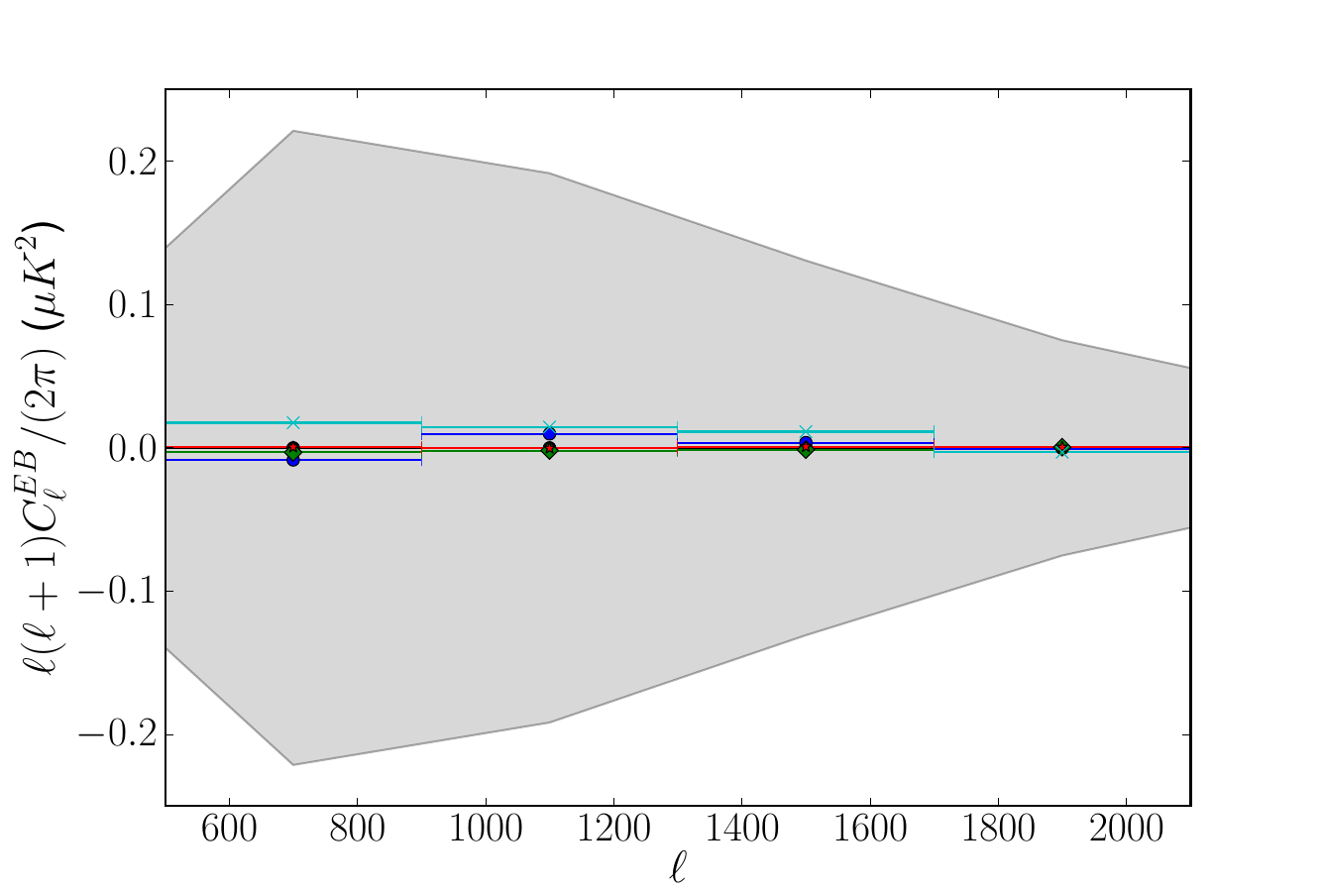}
\end{center}
\caption{\label{fig:total_systematics_EB}
Impact of instrumental systematic uncertainties on the \cleb\ power spectra, as described in \secref~\ref{sec:systematics}. 
See \figref~\ref{fig:total_systematics_BB} for details on the individual sources of uncertainty. 
The shaded region shows the $1\sigma$ boundaries of the uncertainty in the \cleb\ self-calibration procedure described in \secref~\ref{sec:self_calibration}. 
Note that all of the systematic uncertainties are much smaller than this statistical uncertainty.}
\end{figure}

The method outlined above was used to investigate five systematic instrumental effects: uncertainty in instrument polarization angle;  uncertainty in relative pixel polarization angles; uncertainty in instrument boresight pointing model; differential pointing between the two bolometers in a pixel; and relative gain calibration uncertainty between the two bolometers in a pixel. 
All of these instrumental systematic uncertainties have also been described analytically \citep{shimon2008,miller2009}.
All five were found to produce \clbb\ well below the statistical uncertainty in the measurement of \clbb, and \cleb\ substantially smaller than the signal discussed in \secref~\ref{sec:self_calibration}.
The simulated contamination is shown in \figrefs~ \ref{fig:total_systematics_BB} and \ref{fig:total_systematics_EB}. 
The results for \dlbb\ are enumerated in \tabref~\ref{tab:sys-bias}.
Each individual simulation is described in more detail below.

\subsubsection{Uncertainty in polarization angle}
\label{sec:sys-abspol}

All pixel polarization angles are referenced to the instrument's global reference angle. 
How this angle maps to the sky is the sum of different contributions, described in equation \ref{total_polarization_angle}. 
The miscalibration of this angle has been studied analytically by \cite{ksy2013}.
Simulations with an incorrect instrument polarization angle were consistent with these analytic results.
We simulated 100 realizations of miscalibration of the instrument polarization angle, and found that they produced bias in \cleb\ and \clbb\ expected from the analytical calculation.
For the \pb\ \clbb\ results, the global reference angle was measured using \cleb\ with an uncertainty of \rotuncert\textdegree\ as discussed in \secref~\ref{sec:self_calibration}.
All of the relative polarization angle uncertainties were simulated 100 times, and in each simulation the instrument angle was measured from \cleb, with realistic variance in the angle added to the noiseless simulations.
In this way, we calibrated the instrument angle for the systematic uncertainty simulations in the same way it is calibrated in the analysis. 
%The residual bias in \cleb\ and \clbb\ from polarization angle uncertainty after self-calibration is shown in \figrefs~\ref{fig:total_systematics_BB} and \ref{fig:total_systematics_EB}. 

\subsubsection{Uncertainty in relative pixel polarization angle}
\label{sec:sys-relpol}

The relative polarization angles of each pixel are measured using \taua, as described in \secref~\ref{sec:polmodel}.
We simulate the noise in this measurement with two random components: one component which is common across the detectors in each wafer (the uncertainty of wafer-averaged polarization angles) and a component that is a pixel-by-pixel random uncertainty within each wafer (the individual pixel polarization angle uncertainty).
The amplitude of each is based on the measurement uncertainty.
We also include day-to-day variations in the instrument polarization angle at the largest level allowed by measurements of \taua.  
Note that we do not expect or see evidence for day-to-day variations, however this treatment accounts for any possible rotation (jitter) of the stepped and fixed HWP.
The combined uncertainty in \cleb\ and \clbb\ due to polarization angle calibration uncertainty, after self-calibration using \cleb, is shown in \figrefs~\ref{fig:total_systematics_BB} and \ref{fig:total_systematics_EB}. 
We found that the global reference angle uncertainty has a somewhat larger contribution to this uncertainty than the relative pixel polarization angle uncertainty.
The grey-shaded region in \figrefs~\ref{fig:total_systematics_BB} and \ref{fig:total_systematics_EB} shows the 1$\sigma$ bounds on the cumulative bias limit for the polarization angles from 100 realizations after the self-calibration procedure.

\subsubsection{Uncertainty in the reconstructed telescope pointing}
\label{sec:sys-abspoint}

The effect of incorrect pointing reconstruction can be evaluated in this simulation pipeline by scanning the noiseless map into timestreams using one pointing model -- the scanning pointing model -- but then reconstructing the map using a second pointing model -- the mapping pointing model. 
Measuring the spurious \cleb\ and \clbb\ created by this procedure is a measure of how different the pointing models are, referenced to power spectrum space.

The covariance matrix of the five parameters in the pointing model describes the constraints that the pointing data has put on the model parameters. 
For the pointing model to be precise enough, inaccuracies in these parameters within the space of this covariance matrix must be acceptable. 
100 realizations of the pointing model within the parameter covariance matrix were generated and used as the mapping pointing model in the simulation. 
The mean of the spurious signal created in these simulations was found to be negligible compared to the systematic uncertainty in the pointing model described in the next paragraph.

As described in \secref~\ref{sec:pointing}, systematic differences in pointing reconstruction were noticed depending on the sources used to create the pointing model. 
To ensure that these differences were unimportant in this measurement of \clbb, simulations were done with each of these systematically different pointing models used as the mapping pointing model. 
The largest spurious \clbb\ and \cleb\ found in these simulations is shown in \figrefs~\ref{fig:total_systematics_BB} and \ref{fig:total_systematics_EB}. 
This level of systematic error in pointing model would have been responsible for more beam-smearing than was observed in the \pb\ maps (as described in \secref~\ref{sec:beammodel}), but the spurious \cleb\ and \clbb\ are still small compared to the statistical uncertainties in the measurement.
While sufficiently accurate for the measurement of \clbb reported here, in the future, we plan to establish a more precise and consistent pointing model for \pb\ through more detailed pointing calibration observations.

\subsubsection{Differential pointing between two pixel-pair bolometers}
\label{sec:sys-diffpoint}

The differential pointing between two detectors in a pixel is measured from observations of planets. 
It is estimated independently for each HWP position. 
The mean differential pointing magnitude is 5\arcsec.
This is one of the most important instrumental systematic effects because it creates spurious polarization proportional to the derivative of the CMB intensity. 

Data averaged over different angles between the sky polarization and the differential pointing vector acts to average out the effect of differential pointing. 
This averaging out is provided by sky rotation as the CMB patch rises and sets, and HWP rotation from one angle to another \citep{miller2009}.  
The simulations show that a majority of this leakage-mitigation provided by the \pb\ observation strategy has occurred after several days of observation. 
\figrefs~\ref{fig:total_systematics_BB} and \ref{fig:total_systematics_EB} show the spurious \clbb\ and \cleb\ signals created by the differential pointing in an entire season.

\subsubsection{Uncertainty in pixel-pair relative gain}
\label{sec:sys-relgainsim}

Miscalibrations of relative bolometer gains in a pixel pair will ``leak'' temperature signal into $Q$ or $U$. 
A systematic miscalibration between two bolometers does not necessarily lead to a significant systematic bias in polarization maps. 

The relative gain model we use has a term motivated by the polarization of the thermal calibration source which depends on the angle of the HWP, and a term motivated by variations in detector properties which has no HWP dependence. 
Uncertainty in either of these terms can lead to leakage of temperature into polarization. 
We evaluate our uncertainty in the term that changes with HWP position by comparing two different gain models with an independent determination of this term, as described in more detail below.
We evaluate uncertainty in the term that does not change when the HWP rotates in two ways: via a comparison of different gain models with separate measurements of this HWP-independent term, and also via differential gain map-making described in \secref~\ref{sec:sys-special-diffgainmap}. 

The simulation pipeline described above is used to compare relative gain models. 
In each case, a simulated map with no $B$-modes is ``observed,'' producing timestreams using the gain model under question, and then reconstructed using the standard analysis gain model. 
The level of resulting \clbb\ quantifies the difference in these gain models in power spectrum space.

As explained in \secref~\ref{sec:polmodel}, each pixel's polarization angle relative to the instrument frame is measured using \taua. 
This fit also returns a single value for the average relative gain miscalibration over the course of the year,  which we use to independently determine our non-HWP dependent gain model term. 
The difference between this measurement and the measurement by planets that is used to calibrate the pixel-pair relative gain was analyzed with the simulation pipeline.
The resulting \clbb\ bias is shown in \figref~\ref{fig:total_systematics_BB} and enumerated in \tabref~\ref{tab:sys-bias}.

Elevation nods can also be used to establish the relative gain between detectors. 
We use this technique to determine our HWP-dependent relative gain model term and compare to the planet-derived term with the same simulation process. 
We find the difference to be small, as shown in \figrefs~\ref{fig:total_systematics_BB} and \ref{fig:total_systematics_EB} and \tabref~\ref{tab:sys-bias}. 

Our normal procedure to correct for gain drift over the duration of a scan is to interpolate our gains between measurements of the thermal calibration source taken at the beginning and end of hour-long observation periods.
In order to understand the impact of potential errors in this interpolation, we constructed a set of gains based only on the measurements taken at the beginning of every hour and thus use no interpolation.
We find the impact, evaluated through a simulation comparing these two models, to be negligible as shown in \tabref~\ref{tab:sys-bias}. 
All four probes of the relative gain model described here show that the uncertainty in the \pb\ relative gain model is small compared to the statistical uncertainty in \clbb.
We perform a further systematic check on all sources of differential gain via a cross-correlation of temperature maps with $B$-mode maps, described in \secref~\ref{sec:sys-special-tbcorr}, and find consistency with zero leakage to within this test's statistical power. 

\subsubsection{Crosstalk in the multiplexed readout}
\label{sec:crosstalk}

Crosstalk from one bolometer to another due to coupling in the multiplexed readout will lead to polarization and temperature leakage from a point outside the main beam, creating a localized, polarized near side lobe. 
Crosstalk is strongest in bolometer channels that share a SQUID in the frequency-domain multiplexed readout, and are closest together in bias frequency,
as described in \citet{Dobbs2012}. 
\pb\ bolometers of this type show nominal crosstalk of about 1\%. 
Simulations were done with 2\% crosstalk between all nearest-neighbor bolometers; we ran 10 simulations of five typical days of \pb\ observations. 
Because of the frequency schedule in the 8x multiplexed readout, at least half of the pixels will see crosstalk from another pixel that entirely leaks temperature to polarization; in practice the fraction is greater than half because the existence of non-functioning bolometers also creates unbalanced crosstalk. 
All of this means that these simulations should overestimate the effect in \clbb\ of an average 1\% electrical crosstalk. 
The simulation showed temperature to polarization leakage that produces a maximum spurious polarized signal of \dlbb$ < 1.7 \times 10^{-4}$\,\uKsq. 
This simulation was done without simulating stepping of the HWP; observing at several HWP angles mitigates this already small effect.

\subsubsection{Differential beam shapes}
\label{sec:sys-beams}

In \secref~\ref{sec:beammodel}, we described measurements of the individual beam shapes of each detector, which can be used to measure differences in beam shape between the two bolometers measuring orthogonal polarization in one pixel.
These differences can create spurious polarization signals by ``leaking'' temperature into polarization \citep{miller2009}.

The differential beam shape leakage in \pb\ can be modeled as additional terms in the pointing matrix corresponding to leakage from the zero, first, and second derivatives of the temperature map to polarization.  
We fit our elliptical Gaussian beam parameters to a model of derivatives of the beam, and then run simulations to measure the systematic bias in the measured power spectra.  
Differential beam size couples to the zeroth and second derivatives, while differential pointing couples to the first derivative, and differential ellipticity to the second derivative. 
The measured \Pb\ beam and pointing over a representative period of 5 days were used to sample a simulated \lcdm\ realization of the CMB and it's derivative maps.
The leakages into \cleb\ and \clbb, found to be negligible, are shown in \figrefs~\ref{fig:total_systematics_BB} and \ref{fig:total_systematics_EB}.

\subsection{Special analyses focusing on instrumental effects}
\label{sec:sys-special}

Because of the importance of relative gain uncertainty described in \secref~\ref{sec:sys-relgainsim}, we searched for temperature to polarization leakage in the \pb\ data in two ways: differential-gain map-making (\secref~\ref{sec:sys-special-diffgainmap}), and temperature-polarization map correlations (\secref~\ref{sec:sys-special-tbcorr}). We also searched for effects that were correlated with position on the ground due to the large thermal emission from the earth relative to the sky.  We created a scan synchronous signal template and used it to remove these effects (\secref~\ref{sec:sys-ground}).

\subsubsection{Differential gain map making}
\label{sec:sys-special-diffgainmap}
Differential gain maps are constructed by extending the model for polarized timestreams to include a temperature component due to a possible mismatched calibration of the two detectors in a pixel-pair, 

\begin{equation} d_t = G_t + Q_t \cos(2\phi_t) + U_t \sin(2\phi_t). \end{equation}

\noindent We construct maps individually for each wafer's $Q$ and $U$ pixels. 
This is motivated by the idea that the dominant systematic relative gain error could result from differential spectral band pass, which might arise from common fabrication errors within similarly oriented pixels on one wafer. 
Correlation of $G$ maps with temperature maps can be used to estimate differential gain leakage. 
No power is detected in any of the TG cross-spectra; the measured leakage value for each pixel type on each wafer is consistent with zero. 
Simulations that include wafer common mode differential gain leakage at the level constrained by the TG cross spectra show negligible systematic bias.

\subsubsection{Temperature-polarization map correlations}
\label{sec:sys-special-tbcorr}

We investigated the \pb\ maps for unexpected correlations between temperature and polarization.
Specifically, we looked at correlations between $T$ and $B$ maps in 2-dimensional Fourier space. 
Any temperature leakage into $Q$ or $U$ is expected to average out to first order in \cltb .
However, if the Fourier modes are weighted by sine or cosine of the Fourier plane azimuthal angle before performing the azimuthal averaging, the temperature to polarization leakage signal is preserved. 
This weighting scheme has the added benefit of averaging out any $E$ to $B$ leakage due to polarization angle rotation error, making it easier to isolate the effects of gain leakage.

We form estimators for temperature to polarization leakage based on $TB$ using this weighting and compare simulations without gain leakage to our real data.
The signal found in these correlations is consistent with zero-leakage with a minimum PTE of 27\% across the three patches. 
We ran 100 additional simulations at each of two different levels of uniform leakage from temperature to polarization. 100\% of these simulations with a 0.5\% leakage show a larger leakage than seen in the real data, and 86\% of simulations with a 0.3\% leakage show larger values than observed with the real data. These simulations demonstrate the real leakage is less than 0.5\% and likely less than 0.3\%, corresponding to at most 75\% (and likely less than 35\%) of the expected BB power at $\ell=700$.
Note that although these numbers are potentially large, this test is consistent with zero gain leakage and, of the several tests presented, has the least constraining power on gain leakage.

\subsubsection{Scan-synchronous template removal}
\label{sec:sys-ground}

A scan-synchronous template is computed for every 15 minute CES and subtracted from the individual azimuth scans~(\secref~\ref{sec:ana-filtering});
however, some residual contamination that changes in time could remain. 
We make maps for each wafer in ground coordinates and then use these maps to add contamination to the standard CMB signal-only simulations.
Taking the difference between these ground-contaminated simulations and the standard signal-only simulations, applying the scan-synchronous signal filter to both,
we constrain the expected residual power in ground signal not removed by the filter
to be $\ell(\ell+1)C_\ell^{BB}/(2\pi) = 1.7\times 10^{-4}$\,\uKsq\ in the band powers reported here.

We confirmed that the additional absorptive shielding installed above the primary mirror in the middle of the season did in fact eliminate the far side lobe
it was designed to mitigate, reducing the amplitude of the scan synchronous signal by more than two orders of magnitude. 
We apply scan-synchronous signal filtering even after the visor installation to minimize the effect of any residual ground pickup.

\subsection{Null tests}
\label{sec:sys-null}

The \pb\ null test framework is used to show that the data set is internally consistent and to search for possible systematic contamination in the power spectrum.
In a null test, the data set is split into two parts based on configurations
associated with possible sources of contamination or miscalibration.

In general, we cannot construct a map that we expect to have no signal
because maps of different data sets have different coverage and 
cross-linking, and thus different biases due to analysis filtering effects.
Because of this, we construct our null estimator in power spectrum space, which requires correction of the individual transfer functions for the subsets of the data. 
Explicitly, a binned null power spectrum is defined by
\begin{align}
\hat{C}_b^{\rm null} = \sum_{b'\ell}[
(\mathbf{K}_{bb'}^{A})^{-1}\mathbf{P}_{b'\ell}\tilde{C}_\ell^{A} + (\mathbf{K}_{bb'}^{B})^{-1}\mathbf{P}_{b'\ell}\tilde{C}_\ell^{B} \notag \\[-0.7em]
-2 (\mathbf{K}_{bb'}^{AB})^{-1}\mathbf{P}_{b'\ell}\tilde{C}_\ell^{AB} ],
\end{align}
where
\begin{equation}
\mathbf{K}_{bb'}^{i} = \sum_{\ell\ell'}\mathbf{P}_{b\ell}\mathbf{M}_{\ell\ell'}^{AB} F_{\ell'}^{i} B_{\ell'}^2 \mathbf{Q}_{\ell'b'};
\,\,
i\in A, B, AB.
\end{equation}
Here $\mathbf{M}_{\ell\ell'}^{AB}$ is computed analytically from the overlapping sky region between two data sets;
$\tilde{C}_{\ell'}^{A}$ and $\tilde{C}_{\ell'}^{B}$~($F_{\ell'}^{A}$ and $F_{\ell'}^{B}$) correspond to pseudo-power spectra~(transfer functions) for each data set;
$\tilde{C}_{\ell'}^{AB}$~($F_{\ell'}^{AB}$) corresponds to cross-pseudo-power spectra (transfer functions) between two data sets.
We form these spectra and transfer functions by cross-correlating the daily maps in the same fashion as the standard pipeline, but for only the selected data.
We also apply the binning and interpolation operators to evaluate the binned true null spectrum band powers~$\hat{C}_b^{\rm null}$.
We estimate the $EB$ and $BB$ null binned power spectra and check for consistency with the results of 500 \mc\ simulations that include signal and white noise.

\subsubsection{Data splits}

The null tests are performed for several interesting splits of the data, chosen to be sensitive to various sources of systematic contamination or miscalibration.
We also required that the data divisions for different null tests be reasonably independent. The data splits are:

\begin{itemize}

\item ``First half vs.\ second half'': probes time variation on month-long time-scales. This test is sensitive to systematic changes in the calibration, beams, telescope, and detectors, and effects due to the mid-season addition of absorptive shielding above the primary mirror (see \secref~\ref{sec:inst-telescope} and \secref~\ref{sec:sys-ground}).

\item ``Rising vs.\ setting'': checks for systematic bias due to poor sky rotation.
This is also sensitive to residual ground signal via the far sidelobe, which for RA23 sees a nearby hill in only the setting scans~(\secref~\ref{sec:ana-filtering}).

\item ``High elevation vs.\ low elevation'': tests for contamination caused by noise or glitches due to the faster telescope motion required at higher elevation.

\item ``High gain vs.\ low gain'': probes for problems due to linearity or saturation power of the detectors, and checks for miscalibration.

\item ``Good vs.\ bad weather'': checks for residual problems after the PWV cut~(\secref~\ref{sec:cuts}) is made.

\item ``Pixel type'': each detector wafer is fabricated with pixels at two different polarization orientation angles. We split the data into the two individual types of pixels to check for systematic contamination or miscalibration
by different cross-linking, bandwidth, or microfabrication differences.

\item ``Left-side vs.\ right side'': checks for optical distortion on one side of the focal plane versus another, or for different map coverage.

\item ``Left- vs.\ right-going subscans'': probes for residual atmosphere (which is asymmetric in telescope direction due to wind), and for contamination due to vibration, which may be asymmetric in velocity. 

\item ``Moon distance'': checks for residual contamination after setting the moon proximity threshold for an observation to be considered for analysis.

\end{itemize}

We also established a ``Sun distance'' null test, but it was highly correlated with the ``high gain vs.\ low gain'' test for RA4.5, and also correlated with the ``first half vs.\ second half'' test for RA12 and RA23, so we did not include it.
This left nine null tests to analyze as described below.

\subsubsection{Analysis}

For each null power spectrum bin~$b$, we calculate the statistic $\chi_{\rm null}(b)\equiv \hat{C}_b^{\rm null}/\sigma_b$,
where $\sigma_b$ is a \mc\ - based estimation of the corresponding standard deviation, and its square $\chi_{\rm null}^2(b)$.
$\chi_{\rm null}(b)$ is sensitive to systematic biases in the null spectra, while $\chi_{\rm null}^2(b)$ is more sensitive to outlier bins.

\begin{figure}[htbp]
 \centering
 \includegraphics[width=3.2in]{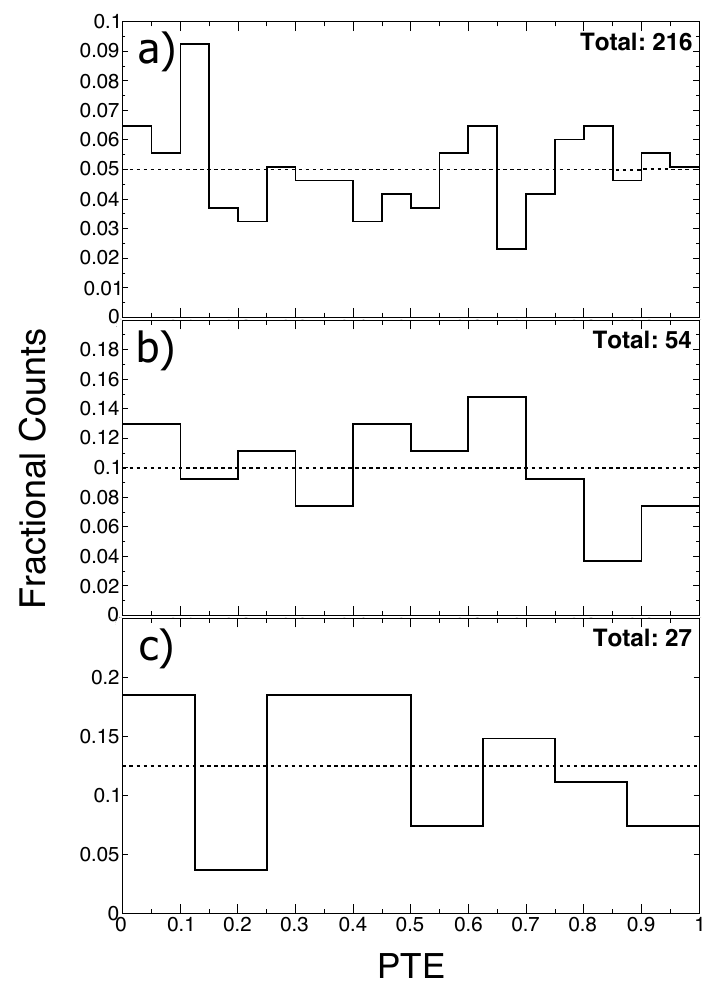}
 \caption{Each plot shows a PTE distribution from the null suite of the $C_b^{BB}$ and $C_b^{EB}$ power spectra of the three patches.
 a), b), and c) corresponds to distribution of $\chi_{\rm null}^2(b)$, $\chi_{\rm null}^2$ by spectrum,
 and $\chi_{\rm null}^2$ by null test, respectively.
 Each is consistent with the uniform expectation.}
 \label{fig:dist_chi2b}
\end{figure}

\begin{table*}
\begin{center}
\caption{\label{tab:pte_summary}PTEs resulting from the null test framework. 
No significantly low or high PTE values are found, consistent with a lack of  systematic contamination or miscalibration in the \pb\ data set and analysis. 
Note that the PTE values in each patch are not independent from each other. }
\begin{tabular}{cccccc}
\tableline
\tableline
       & average of & extreme of & extreme of & extreme of & total \\
 Patch & $\chi_{\rm null}(b)$ & $\chi^2_{\rm null}(b)$ & $\chi^2_{\rm null}$ by {\it EB}/{\it BB} & $\chi^2_{\rm null}$ by test & $\chi^2_{\rm  null}$\\
\tableline
 RA4.5 & 11.6\% & 16.6\% & 20.6\%           & 21.8\%           & 14.0\% \\
 RA12  & 92.4\% & 84.2\% & 60.8\%           & 23.8\%           & 52.6\% \\
 RA23  & 75.2\% & 61.6\% & \phantom{0}6.0\% & \phantom{0}7.0\% & 18.6\% \\
\tableline
\end{tabular}

\end{center}

\end{table*}

To probe for systematic contamination that is focused in a particular power spectrum or null test data split, we calculate the sum of $\chi_{\rm null}^2(b)$ over $500 < b < 2100$ for $EB$ and $BB$ separately, (``$\chi^2_{\rm null}$ by spectrum''), and the sum of both these spectra for a specific test (``$\chi^2_{\rm null}$ by test'').
\figref~\ref{fig:dist_chi2b} shows the PTE distribution of the $\chi^2_{\rm null}$ by (a) bin, (b) spectrum, and (c) test for the three patches. 
We require that each of these sets of PTEs each be consistent with a uniform distribution, as evaluated using a KS test, requiring a p-value (probability of seeing deviation from uniformity greater than that which is observed given the hypothesis of uniformity) greater than 5\%.
These distributions are consistent with a uniform distribution from zero to one.

We create test statistics based on these quantities to search for different manifestations of systematic contamination. 
The five test statistics are (1)~the average value of $\chi_{\rm null}$;
the extreme value of $\chi^2_{\rm null}$ by (2)~bin, (3)~spectrum, and (4)~test; 
and (5)~the total $\chi^2_{\rm null}$ by summing up the nine null tests. 
In each case, the result from the data is compared to the result from simulation, and PTEs are calculated.
Finally, we combine each of the test statistics, and calculate the PTE of that final test statistic, requiring it to be greater than 5\%. 
\tabref~\ref{tab:pte_summary} shows summary of the PTE values of each test statistic for each patch.
Comparing the most significant outlier from the five test statistics with that from simulations,
we get PTEs of 32.8\%, 55.6\%, and 18.0\% for RA4.5, RA12, and RA23 respectively.
We achieve the requirements described above, finding no evidence for systematic contamination or miscalibration in the \pb\ data set and analysis.

\subsection{Cross-check using a second pipeline}
\label{sec:ubpipe}

Concurrently, we have been developing an alternate data processing pipeline that was used to cross-check the results presented here. 
Its full description will be given in a forthcoming publication; here we highlight its most salient features. 

In the time domain, the alternate pipeline applies the same filters as the primary pipeline, but corrects for them while estimating the sky signals as part of the map-making procedure, following \citet{Stompor2002}. 
The recovered maps provide unbiased renditions of the sky signal, with the filtered modes effectively marginalized over. 
This is numerically challenging so we use a divide-and-conquer approach, which results in unbiased but slightly sub-optimal maps. 
The maps are estimated in the HEALPix pixelization \citep{Gorski2005} with $N_\textrm{side} = 2048$, so no flat-sky assumption is adopted. 
We produce the maps of three Stokes parameters and the $Q$ and $U$ maps are used to estimate the polarized power spectra of the sky signals. 
This is done with power spectrum estimation software packages based either on the pure-pseudospectra \citep{Smith2006}, \textsc{xpure} and \textsc{x$^2$pure} \citep{xpure,Grain2012,ferte2013}, or the standard pseudospectra \textsc{xpol} \citep{tristram2005} approaches. 
The mode-coupling matrices are computed explicitly by directly summing the required Wigner-3j symbols based on the geometry of the observed patches, noise weights and apodizations. 
The final spectra are calculated as weighted averages of the cross-spectra of 8 maps made of disjoint subsets of all daily maps, and \mc\ simulations are employed to estimate the final uncertainties of the computed spectra. 
The results of this alternate pipeline are consistent with the results of the primary pipeline described in this publication.

\begin{figure*}[htpb]
 \centering
 \includegraphics[width=7in]{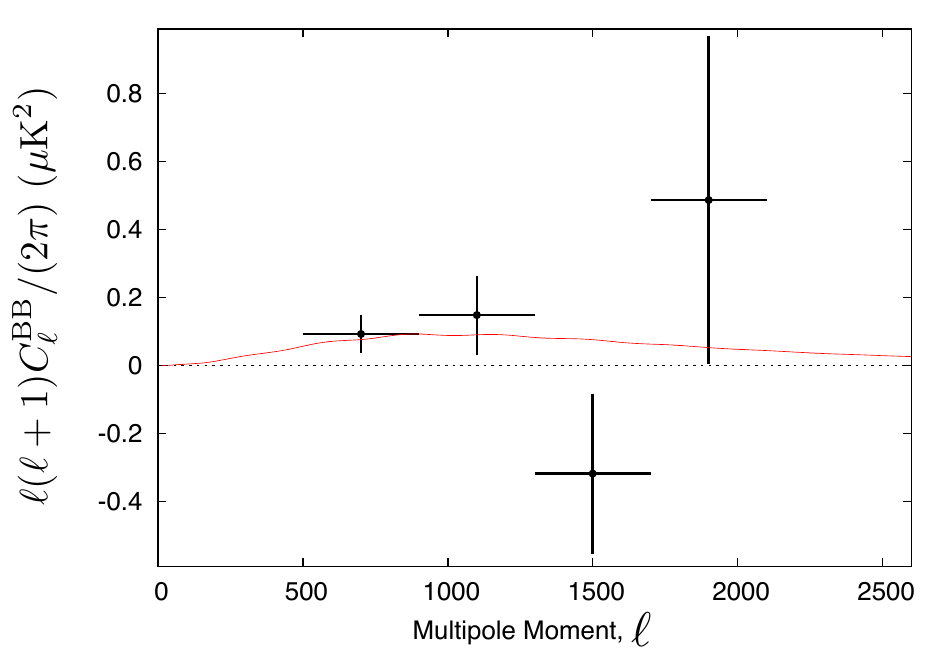}
 \caption{\label{fig:resultBB}Binned \clbb\ spectrum measured using data from all three patches~($\sim$25\,\sqdeg).
 A theoretical \wmap-9 \lcdm\ high-resolution \clbb\ spectrum with \ABB$ = 1$ is shown.
 The uncertainty shown for the band powers is the diagonal of the band power covariance matrix, including beam covariance.}
 \label{fig:bmode_spectrum}
\end{figure*}

\subsection{Blind analysis}
\label{sec:blind}

The possibility of data analyzers biasing their result toward their own preconceptions, known as ``observer bias'', is a form of systematic bias that can affect the result of an experiment \citep{Klein:2005di}.
Examples of preconceptions include theoretical predictions, the statistical significance that the team expect to obtain, or consistency with previous measurements. 
Since it is difficult to estimate the effects of observer bias, we employed an analysis methodology designed to minimize its impact. 

We have adopted a blind-analysis framework, which is a standard technique to
minimize observer bias. %\citep{Klein:2005di}.
In our framework, no one in the team viewed the measured \clbb\ values, the deflection power spectra based on $B$-modes \citep{PB_CLdd_2014,PB_GalaxyCross_2014}, or the corresponding maps, until we eliminated possible sources of observer bias by finalizing calibration, filtering, data selection, data validation and showed that all systematic uncertainties were small.
This framework forced us to develop quantitative tools, including null tests and simulations, that convincingly argued for analysis choices and constraints without showing \clbb, thus removing the possibility that people within the team would be more convinced by an argument or method because of the \clbb\ that it produced.
Other power spectra and maps were used as subsidiary information in this work, and they were unblinded in stages during the analysis procedure.

In fact, after un-blinding \clbb, questions came up about how well we had constrained electrical crosstalk, and how robust our estimate of the binned power spectrum uncertainty was. 
Finding our previous argument constraining electrical crosstalk weak, we developed the simulation shown in \secref~\ref{sec:crosstalk}, where we estimated that electrical crosstalk is one of our smallest systematic uncertainties. 
Investigating our binned power spectrum uncertainties, because of comparisons with a second pipeline, we found an error in our uncertainty estimation code. 
This was an error that could have been found while we were blind, but it was not. 
The error did not affect the central values of the measurement.
We corrected this error, resulting in a reduction in the significance of our measurement by about 18\% between un-blinding and the results presented here. 
The qualitative consistency of the measurement with theory was not changed, the change was motivated by a pipeline comparison, and it reduced the significance of our measurement; we do not believe that this was a significant opportunity for the result to be incorrectly affected by observer bias.

%%%%%%%%%%%%%%%%%%%%
\section{Power spectrum results}
\label{sec:results}

A single estimate of the \clbb\ power spectrum from the three patches is created using the individual patch band powers and their covariance matrices. 
This \clbb\ spectrum is shown in \figref~\ref{fig:resultBB}.
We calculate the PTE of these band powers to the \wmap-9 \lcdm\ \clbb\ spectrum; including statistical uncertainty and beam covariance, this PTE is \ptelcdm. 
\Tabref~\ref{tab:bandpowers} enumerates the band powers reported here.

\begin{table}
\begin{center}
\caption{\label{tab:bandpowers}Reported \pb\ band powers and the diagonal elements of their covariance matrix}
\begin{tabular}{ccc}
\hline
\hline
Central~$\ell$ & \dlbb~[\uKsq] & $\Delta$\{\dlbb\}~[\uKsq] \\
\hline
\phantom{0}700 & $\phantom{+}0.093$ & $0.056$ \\
          1100 & $\phantom{+}0.149$ & $0.117$ \\
          1500 & $-0.317$           & $0.236$ \\
          1900 & $\phantom{+}0.487$ & $0.482$ \\
\hline
\end{tabular}
\end{center}
\end{table}

We fit the band powers to a \lcdm\ cosmological model with a single \ABB\ amplitude parameter. 
We find \ABBmeas, where $A_{BB} = 1$ is defined by the \wmap-9 \lcdm\ spectrum. 
To calculate the lower bound on the additive uncertainties on this number, we linearly add, in each band, the upper bound band powers of all the additive systematic effects discussed in \secref~\ref{sec:systematics}, and the uncertainty in the removal of $E$ to $B$ leakage. 
We then subtract this possible bias from the measured band powers, and calculate \ABB. 
This produces a lower \ABB, and sets the lower bound of the additive uncertainty. We then repeat the process to measure the upper bound.
The multiplicative uncertainties are the quadrature sum of all the multiplicative uncertainties discussed in \secref~\ref{sec:systematics}.
\Tabref~\ref{tab:summary_A_BB} summarizes all the systematic uncertainties in the measurement of \ABB.

The measurement rejects the hypothesis of no \clbb\ from lensing with a confidence of \rejectnull. 
This is calculated using the bias-subtracted band powers described above (the most conservative values to use for rejecting this null hypothesis), and integrating the likelihood of \ABB$ > 0$.

\begin{table*}
 \begin{center}
  \caption{\label{tab:summary_A_BB}Summary of possible contributions to the amplitude \ABB\ from major sources of systematic uncertainty.}
  \begin{tabular}{p{1.25in} p{3.1in} p{0.74in}}
    \hline
    \hline
    \centering Type &\centering Source of systematics & Effect on \ABB \\
    \hline
    Systematic uncertainty:
    & Galactic dust & $0.045$ \\
    astrophysical foreground 
    &	Galactic synchrotron & $0.003$\\
    (\secref~\ref{sec:fg}) &	Radio galaxies & $0.011$\\
    &	Dusty galaxies & $0.004$\\
    \cline{2-3} 
    Systematic uncertainty:
    &	Differential \& boresight pointing~(\secref~\ref{sec:sys-abspoint}, \ref{sec:sys-diffpoint})  & $0.017$\\ 
    instrument 
    &	Instrument \& relative polarization angle~(\secref~\ref{sec:sys-abspol}, \ref{sec:sys-relpol}) & $0.014$\\
    &	Pixel-pair relative gain: HWP-independent~(\secref~\ref{sec:sys-relgainsim}) & $0.002$\\
    & 	Pixel-pair relative gain: HWP-dependent~(\secref~\ref{sec:sys-relgainsim}) & $0.010$\\
    &	Pixel-pair relative gain: drift~(\secref~\ref{sec:sys-relgainsim}) & $0.001$\\
    &	Differential beam ellipticity~(\secref~\ref{sec:sys-beams}) & $0.001$\\
    &	Differential beam size~(\secref~\ref{sec:sys-beams}) & $0.003$\\
    &	Electrical crosstalk~(\secref~\ref{sec:crosstalk}) & $0.002$\\
    \cline{2-3}
    Systematic uncertainty:
    &      Scan synchronous template~(\secref~\ref{sec:sys-ground}) & $0.002$\\
    analysis
    &        $E$-to-$B$ leakage subtraction~(\secref~\ref{sec:mll_and_fl}) & $0.006\pm 0.037$\\
    \hline
    & Total & $0.121\pm 0.037$\\
    \tableline
    \tableline
    Multiplicative effect
    & Statistical variance and beam co-variance~(\secref~\ref{sec:polmodel}) & $\pm 0.041$ \\ 
    & Polarization efficiency~(\secref~\ref{sec:polmodel}) & $\pm0.036$ \\
    & Transfer function~(\secref~\ref{sec:systematics}) & $\pm 0.039$ \\
    \hline
    & Total & $\pm0.067$\\
    \hline
    \hline
    &&\\
  \end{tabular}
 \end{center}
\end{table*}

\section{Summary \& discussion}
\label{sec:summary}

We have reported a measurement of the CMB's $B$-mode angular power spectrum, \clbb, over the  multipole range $500 < \ell < 2100$.  
This measurement is enabled by the unprecedented combination of high angular resolution (3.5\arcmin) and low noise that characterizes the \pb\ CMB polarization observations. 

To validate the \pb\ measurement of this faint signal, we performed extensive tests for systematic errors.  We evaluated nine null tests and estimated twelve sources of instrumental contamination using a detailed instrument model, and found that all the systematic uncertainties were small compared to the statistical uncertainty in the measurement. 
To motivate comprehensive evaluation of the data set and prevent observer bias in data selection and analysis, the analysis was performed blind to the \clbb\ signal; all data selection and analysis choices were fixed and all systematic error tests were completed before any team members looked at the $B$-mode power spectrum. 

\pb\ has reached an important CMB polarization milestone, with noise levels sufficiently low to allow reconstruction of the lensing signal with more precision from polarization than from CMB temperature \citep{HuOkamoto2002}.
We previously presented evidence for gravitational lensing of the CMB in \pb\ data using the non-Gaussianity imprinted in the CMB by LSS \citep{PB_GalaxyCross_2014,PB_CLdd_2014}. 
Those analyses, arising from the same area of sky, are also consistent with the \lcdm\ expectation and give no evidence for significant systematic errors. 
We can calculate the combined significance with which those measurements of non-Gaussian $B$-modes and the \clbb\ measurements reported here reject the hypothesis that there are no CMB lensing $B$-modes. 
In this null hypothesis, the signals are uncorrelated (when using a realization-dependent lensing bias subtraction to calculate the deflection field), so a simple quadrature sum of the rejection significance is appropriate. 
This calculation results in a rejection of the hypothesis that there are no lensing $B$-modes with \rejectnullcombinedsigma\ confidence for a normal distribution.

CMB $B$-mode polarization is emerging as a key observable in modern cosmology. 
Over the next few years, measurements of CMB $B$-mode polarization will allow us to probe LSS in detail to provide insight into fundamental physics, cosmology, and extragalactic astrophysics. 
Detailed analysis of the signal produced by LSS will enable precision characterization of the possible underlying \clbb spectrum from cosmic inflation. 
The measurement of LSS-induced $B$-mode power in \pb\ data, characterized by both its non-Gaussian signature and its \clbb\ power, represents an important step in the rapidly progressing field of CMB $B$-mode science.

\acknowledgments
Calculations were performed on the Central Computing System, owned and operated by the Computing Research Center at KEK, and the National Energy Research Scientific Computing Center, which is supported by the Department of Energy under Contract No. DE-AC02-05CH11231. 
The \Pb{} project is funded by the National Science Foundation under grants AST-0618398 and AST-1212230.
The KEK authors were supported by MEXT KAKENHI Grant Number 21111002, and acknowledge support from KEK Cryogenics Science Center. 
The McGill authors acknowledge funding from the Natural Sciences and Engineering Research Council of Canada, the Canada Research Chairs Program, and Canadian Institute for Advanced Research. 
BDS acknowledges support from the Miller Institute for Basic Research in Science, NM acknowledges support from the NASA Postdoctoral Program, and KA acknowledges support from the Simons Foundation. MS gratefully acknowledges support from Joan and Irwin Jacobs.
All silicon wafer-based technology for \Pb{} was fabricated at the UC Berkeley Nanolab. 
We are indebted to our Chilean team members, Nolberto Oyarce and Jos\'e Cortes. 
The James Ax Observatory operates in the Parque Astron\'{o}mico
Atacama in Northern Chile under the auspices of the Comisi\'{o}n Nacional de Investigaci\'{o}n Cient\'{i}fica y Tecnol\'{o}gica de Chile (CONICYT).
Finally, we would like to acknowledge the tremendous contributions by Huan Tran to the \pb\ project.

\bibliographystyle{apj}

\bibliography{pb1_2013A}

\end{document}